\documentclass[12pt,preprint]{aastex}

\newcommand{\degree}{\mbox{$^{\circ}$}}

\lefthead{Allers et al.}
\righthead{H$_2$ in the Orion Bar}
\begin{document}

\title{H$_2$ Pure Rotational Lines in the Orion Bar}

\author {K. N. Allers\altaffilmark{1}, D. T. Jaffe\altaffilmark{1}, J. H. Lacy\altaffilmark{1}}
\affil{Department of Astronomy,
University of Texas at Austin, Austin, TX 78712-0259}
\author {B. T. Draine}
\affil{Princeton University Observatory, Peyton Hall, Princeton, NJ 08544
\\ 
and Institute for Advanced Study, Einstein Drive, Princeton, NJ 08540
}
\and
\author {M. J. Richter\altaffilmark{1}}
\affil{Department of Physics, University of California, Davis,
1 Shields Avenue, Davis, CA 95616}
\altaffiltext{1}{Visiting Astronomer at the Infrared Telescope Facility, which is operated by the University of Hawaii under Cooperative Agreement no. NCC 5-538 with the National Aeronautics and Space Administration, Office of Space Science, Planetary Astronomy Program.}
% Notice that each of these authors has alternate affiliations, which 
% are identified by the \altaffilmark after each name.  The actual alternate 
% affiliation information is typeset in footnotes at the bottom of the 
% first page, and the text itself is specified in \altaffiltext commands. 
% There is a separate \altaffiltext for each alternate affiliation 
% indicated above. 

% The abstract environment prints out the receipt and acceptance dates 
% if they are relevant for the journal style.  For the aasms style, they 
% will print out as horizontal rules for the editorial staff to type 
% on, so long as the author does not include \received and \accepted 
% commands.  This should not be done, since \received and \accepted dates 
% are not known to the author. 
 
\begin{abstract}
Photodissociation regions, where UV radiation dominates the energetics and  
chemistry of the neutral gas, contain most of the mass in the dense  
interstellar medium of our galaxy.  Observations of H$_2$ rotational and  
ro-vibrational lines reveal that PDRs contain unexpectedly large amounts of  
very warm (400-700 K) molecular gas.  Theoretical models have difficulty explaining the existence of so much warm gas.  Possible  
problems include errors in the heating and cooling functions or in the  
formation rate for H$_2$.  To date, observations of H$_2$ rotational lines  
smear out the structure of the PDR.  
Only by resolving the hottest layers of H$_2$  
can one test the predictions and assumptions of current models. 

Using the Texas Echelon Cross Echelle Spectrograph (TEXES) we mapped 
emission in the H$_2$  
v = 0-0~S(1) and S(2) lines toward the Orion Bar PDR at 2\arcsec \  
resolution. We also observed H$_2$ v = 0-0~S(4) at selected points toward  
the front of the PDR.  Our maps cover a 12\arcsec \ by 40\arcsec \ region of  
the bar where H$_2$ ro-vibrational lines are bright. 
The distributions of H$_2$ 0-0~S(1), 0-0~S(2), and 1-0~S(1) line emission
agree in remarkable detail.

The high spatial resolution (0.002 pc) of our observations allows us to probe  
the distribution of warm gas in the Orion Bar to a distance  
approaching the scale length for FUV photon absorption.  
We use these new observational results
to set parameters for the PDR models described in a companion paper 
(Draine et al.
2005, in prep).  The best-fit model can account for the separation of the H$_2$ emission
from the ionization front and the intensities of the ground state rotational
lines as well as the 1-0~S(1) and 2-1~S(1) lines.  This 
model requires significant adjustments to the commonly used values for the
dust UV attenuation cross section and the photoelectric heating rate.
 
\end{abstract} 
 
% The different journals have different requirements for keywords.  The 
% keywords.apj file, found on aas.org in the pubs/aastex-misc directory,  
% contains a list of keywords used with the ApJ and Letters.  These are  
% usually assigned by the editor, but authors may include them in their  
% manuscripts if they wish.  
 

%\keywords{globular clusters,peanut clusters,bosons,bozos}

% That's it for the front matter.  On to the main body of the paper.
% We'll only put in tutorial remarks at the beginning of each section
% so you can see entire sections together.

% In the first two sections, you should notice the use of the LaTeX \cite
% command to identify citations.  The citations are tied to the
% reference list via symbolic KEYs.  We have chosen the first three
% characters of the first author's name plus the last two numeral of the
% year of publication.  The corresponding reference has a \bibitem
% command in the reference list below.
%
% Please see the AASTeX manual for a more complete discussion on how to make
% \cite-\bibitem work for you.   
\section{Introduction}

A substantial fraction of the dense interstellar medium resides in
clouds where far-ultraviolet photons emitted by hot stars 
dominate both the energetics and the chemistry of the primarily
neutral gas \citep{hollenbach}.  In the dense ISM, the 
neutral photodissociation region (PDR) material includes both
extended molecular
clouds with only modest column densities
(N$\mathrm{_{H_2}}<$10$^{22}
%btd start
{\,\rm cm}^{-2}
%btd end
$) \citep{plume,jansen}
and surface layers of clumps within
higher column density molecular cores \citep{stutzki}.
Within photodissociation regions, the material makes
the transition from hot, ionized gas to cold, molecular 
gas as attenuation of the far-ultraviolet field increases farther from the 
cloud surface.  
As one moves from the ionization front deeper into the molecular cloud,
hydrogen goes from atomic to molecular form and carbon goes from C$^+$
to C$^{\rm o}$ 
and then to CO \citep{th85b,black}.  The rich chemistry of the photodissociated
gas differs significantly from classical dark-cloud chemistry driven by
cosmic-ray ionization \citep{sternberg95}.

The thermal balance in the photodissociation region is intimately 
connected with the radiative transfer for UV photons that drive the
chemistry and energetics and with the chemical state of the material
in different layers of the structure \citep{draineiso}.
The most important heating mechanisms include ejection of photoelectrons
from dust grains \citep[][and refs therein]{bakes,weingartner+draine_2001} 
and the
collisional deexcitation of H$_2$ molecules initially excited by UV
photons \citep{sternberg89}.  Deeper into the regions, 
gas-grain collisions may also heat the gas.  Fine-structure lines of neutral
atoms or of singly ionized species with low ionization potential provide
the cooling in the outer layers of the PDR while CO rotational lines
cool the predominantly molecular inner regions.  Quadrupole rotational
lines of H$_2$ can contribute to the cooling at intermediate depths.

The cooling of PDRs produces a broad variety of line and dust feature
emission, each arising in a particular layer of the photodissociated
structure and each with its own dependence on the density of the region
and on the strength of the incident radiation field.
From the outer, predominantly atomic layers, one sees emission in the 
far-IR fine-structure lines of [CII] (158 $\mu$m) and [OI] (63 and 145 $\mu$m),
as well as lines of [FeII] and [SiII] at somewhat shorter wavelengths. 
The 3.3 $\mu$m feature attributed to PAHs
also appears to arise in this
zone.  Deeper into the cloud, carbon becomes neutral and the 370$\mu$m
and 609 $\mu$m [CI] fine-structure lines become important emitters.  
Farther into the neutral zone, emission from a few molecules, notably
CN and HCO$^+$, is significantly enhanced over dark-cloud values.  Because
of the enhanced temperatures, high-J lines of CO can also trace the 
distribution of PDR material.

The rotational and ro-vibrational transitions of molecular hydrogen are 
particularly useful tracers of the properties of PDRs.  In the outer portions
of the PDR, H$_2$ formed on dust grains is excited by UV photons.   
A radiative cascade through the ro-vibrational
levels of the ground electronic state follows this fluorescent excitation
and produces a distinctive pattern of emergent line strengths
\citep{black, hasegawa}.  At densities greater
than $\sim 5 \times$10$^4$ cm$^{-3}$, collisional de-excitation modifies
the line ratios as it heats the gas \citep{sternberg89,mluhman}.  Models show 
that self shielding allows 
hydrogen to make a transition from predominantly atomic to predominantly
molecular form at a depth into the PDR where the temperature can be
high enough to produce significant populations up to $J=6-8$ 
in the H$_2$ ground rotational state \citep{draineapj}.
This warm molecular gas then produces strong emission in the mid-IR
rotational lines of H$_2$ and may even contribute significantly to
emission in transitions arising from low J levels of the
first excited vibrational state.  The low critical density of the
ground state rotational transitions makes ratios of mid-IR lines
good probes of the temperature in the layers where they arise.

Unfortunately, the line emission from PDRs rarely arranges itself into
the neat stratified pattern one might expect from a homogeneous, 
plane-parallel region.
Furthermore, the extent of emission tracing the surface layers of
clouds, particularly CII and H$_2$ emission, is often larger than one
would naively expect based on likely column densities and dust opacities.
The usual explanation invoked to resolve this discrepancy is that the
PDR material actually resides on the surface layers of clumps and that
UV radiation reaches these surfaces by propagating through a much
more tenuous interclump medium \citep{stutzki,howe,meixner,spaans}.

The first observations of PDRs in the H$_2$ ground-state rotational lines
produced a surprising result: the bulk of the H$_2$
emission was coming from very warm gas.  Observations of the 0-0~S(1)
and S(2) emission from the Orion Bar implied temperatures of 400-1000 K
\citep{parmar}.  Lower spectral and spatial resolution ISO observations of
PDRs in S140, NGC 7023, and NGC 2023 in a larger number of H$_2$
rotational transitions are consistent with temperatures in a similar
range \citep{timmermann,fuente,bertoldiiau}.
These observations have served as a testbed for models of the thermal 
balance and chemistry of PDRs \citep{draineiso,bertoldiiau}.  The 
large number of transitions detected in the ISO observations help 
constrain the models.  The poor spatial and spectral resolution of
these observations, however, leave a number of questions open.  
The width and velocity of the H$_2$ lines cannot be measured 
accurately enough to compare them
to CO, CS, or HCN emission from farther into the PDRs.  The large
beams inevitably lead to an averaging of flux from different layers
within the PDR structure.  This averaging could destroy important
information about the physics and structure of the PDR.

\begin{figure}
\epsscale{1.0}
\plotone{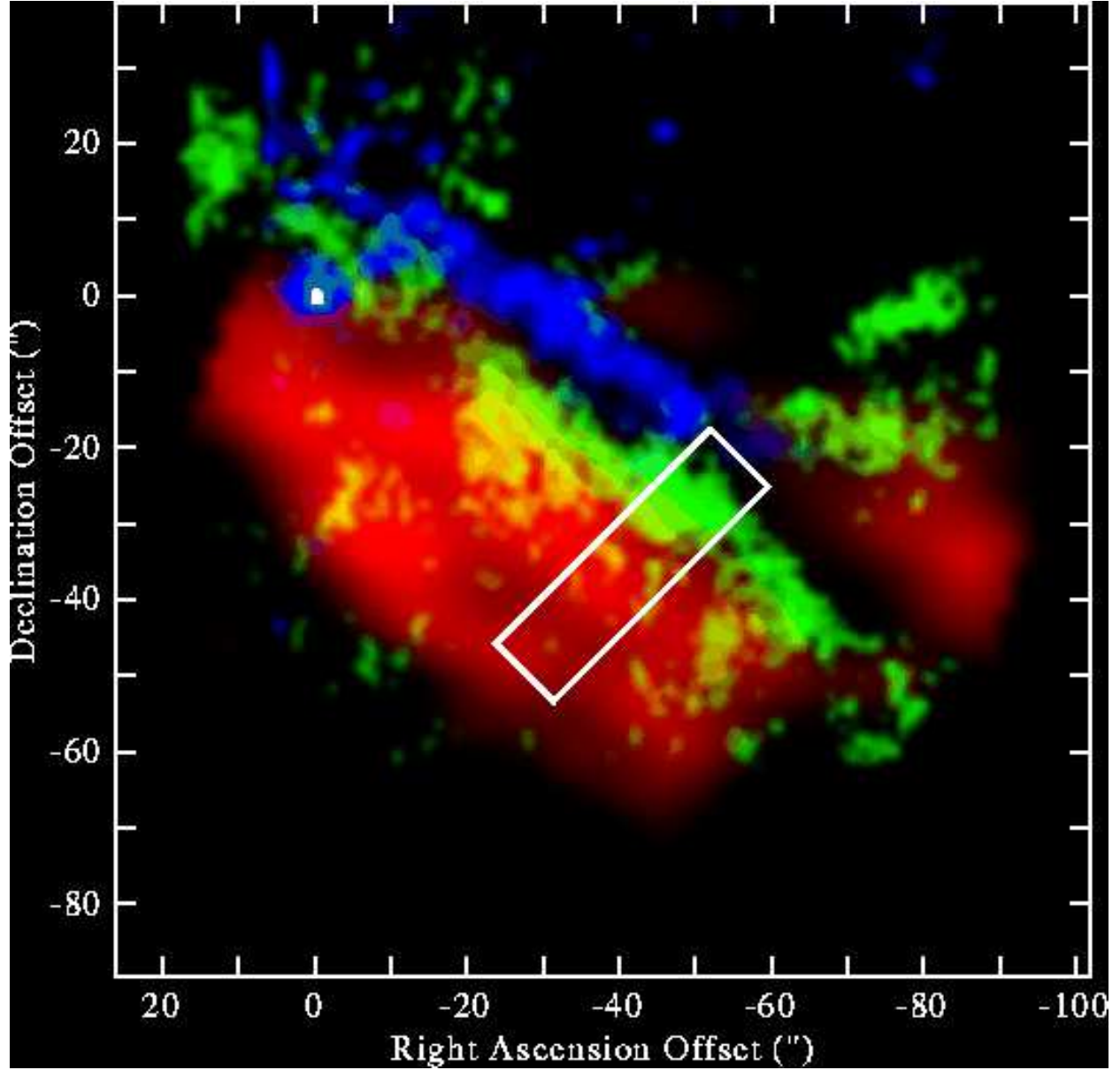}
\caption[1]
%>>>> use \label inside caption to get Fig. number with \ref{}
{\label{context} Distribution of line and dust feature emission from the
Orion Bar \citep{tielens} with the area of our scan
superimposed.  Blue is 3.3 $\mu$m PAH emission, green is 1-0~S(1) line
emission, and red is CO J=1-0 emission.  $\theta^2$A Ori 
(R.A. = 5$^h$ 35$^m$ 22.5$^s$, Dec. = -5\degree\ 24\arcmin\ 
57.8\arcsec\ J2000.0) lies at the (0,0) position of the image.
The top of our maps in Figure \ref{maps} corresponds to the northwest  
edge of the scanned area.
}
\end{figure}

The PDR known as the Orion Bar lies $\sim$2\arcmin\ southeast of the
Trapezium stars in Orion at the interface of the HII region formed by the
Trapezium and dense gas associated with the Orion Molecular Cloud.
The nearly edge-on geometry of the Orion Bar \citep{marconi} lends itself well
to the study of PDRs.  Because of its favorable geometry and relatively
close distance \citep[$\sim$450 pc,][]{hoogerwerf}, it is possible to
observe stratification in the Orion Bar (Figure \ref{context}).  
The ionization front defined by the sharp edge of the radio
continuum emission \citep{felli} lies immediately to the northwest of
the maximum emission in the 3.3 $\mu$m PAH feature \citep{bregman}.
Approximately 15\arcsec\ southeast of the ionization front,
there is a strong maximum in the distribution of H$_2$ 1-0~S(1) line
flux \citep{vanderwerf}.  The peak of the high column density molecular
ridge, as seen in CO, lies an additional 10\arcsec\ to the southeast
\citep[see Fig.1,][]{tielens}.
In this work, we present new maps of the 0-0~S(1) and 0-0~S(2) transitions
of H$_2$ and selected observations of the 0-0~S(4) line toward the 
Orion Bar.  These observations have both high spatial (2\arcsec) and
spectral ($\sim$4 km s$^{-1}$) resolution.  We compare these results
to observations of the 1-0~S(1) line at similar spatial resolution
\citep{vanderwerf}.  The high spatial resolution probes a 
critical scale in Orion (A$_{\rm V}$=1 at 
n$_{\rm H}$=3$\times$10$^4$~cm$^{-3}$ corresponds to 10\arcsec\ 
at the distance of Orion) and the spectral
resolution of our observations is sufficient to resolve lines with the
same width as the CO lines arising in the bar.

By obtaining high spatial resolution H$_2$ pure-rotational line maps, in a
region with nearly edge-on geometry,
we hope to resolve the thermal and chemical structure of
PDRs. We will look not only at the temperature distribution in the Orion Bar,
but also the column densities of H$_2$ and compare the distribution of 
H$_2$ rotational emission to other tracers of gas in PDRs.  We will
use the data to test PDR models of the warm region at the H/H$_2$
interface.

\section{Observations and Data Reduction}

We mapped the H$_2$ v = 0-0~S(1), S(2), and S(4) lines at 17.03~$\mu$m, 
12.28~$\mu$m, and 8.03~$\mu$m (Table \ref{lines}) 
toward the Orion 
Bar in 2002 December. We made the observations using the Texas Echelon 
Cross Echelle Spectrograph \citep[TEXES,][]{lacy} on the 3m NASA Infrared 
Telescope Facility (IRTF).  The spatial scale 
on the 256$^2$ Si:As array was 0.35\arcsec\ 
pixel$^{-1}$.  The spectral resolution of TEXES was determined from 
emission lines (C$_2$H$_2$ and C$_2$H$_6$) toward Titan, assumed to be 
unresolved, observed during the same observing run.   
Table \ref{observations} lists the slit width, length, resolving power, and 
total integration time for each line. 
We oriented the slit parallel to the Orion Bar ionization 
front, at a position angle of 45\degree\ or northeast-southwest. 

With TEXES, the standard method for producing maps of spectral lines
is to step the slit across the object and a portion of adjacent sky, taking a
spectrum at each position, 
with the telescope secondary mirror held fixed (i.e.,
without chopping).  For the S(1) and S(2) lines, 
we mapped the Orion Bar by stepping the telescope from northwest to southeast 
in 1/2 slit width steps (0.7\arcsec\ for S(2) and 1\arcsec\ for S(1)) to 
create 40\arcsec\ long scans. The 0,0 position for the maps is at
R.A. = 5$^h$ 35$^m$ 19.7$^s$, Dec. = -5\degree\ 25\arcmin\ 
28.3\arcsec\ (J2000.0) and the mapped region runs from 13\arcsec\ 
northwest to 27\arcsec\ southeast of this position (Figure \ref{maps}). 
Spectra taken at the end of each scan, where no line
emission is present, are then used as sky frames in the data 
reduction \citep{lacy}.  In Orion, we used the 
last 7\arcsec\ of our scans, the positions  
farthest from the ionization front, for sky subtraction. In order to cover 
the same area with the shorter S(2) slit as we did for the S(1) line, 
we mapped the Bar at 12.28 $\mu$m by making two scans offset by 
$\pm$2\arcsec\ northeast-southwest with respect to the 
center of our 0-0~S(1) map. 
Since autoguiding is unavailable while using TEXES in scan mode, we made 
note of the telescope drift during the scan as
the guide star passed through the boresight at the (0,0) position, and 
recentered the telescope on the boresight before each set of four scans.

We determined our absolute positional uncertainty by scanning the 
Becklin-Neugebauer object (BN) multiple times during our observations.  We 
found that the maximum position drift for the step scans
was 1.3\arcsec\ over the duration of our observations.  
The relative positional uncertainty of our observations 
was determined from the offset of the guide star observed during each scan.
The RMS drift was 0.7\arcsec\ with a maximum of 2.1\arcsec. The main effect 
of these drifts would have been to smear out our beam. 
However, by noting these drifts as they occurred, we were able to correct
for them during the summation that produced the final maps, thereby making 
the smearing negligible compared to our final 2\arcsec\ to 
2.5\arcsec\ resolution.  We therefore estimate the relative 
positional uncertainty in our 0-0~S(1) and S(2)
maps to be less than 0.5\arcsec.

For the S(4) line, we took spectra at 6 different depths into the bar spaced 
1.4\arcsec\ apart.  The slit positions for these observations are superposed
on the map in Figure \ref{areas}. 
We performed sky subtraction by nodding the telescope
every 8-16 seconds between the source position and a reference position 
80\arcsec\ south.

We reduced the raw images of cross-dispersed spectra using the standard
TEXES pipeline reduction program \citep{lacy}.  This program
removes artifacts, flat-fields the frames using images of ambient and
sky loads, extracts the spectra and removes telluric absorption.
The telluric lines (principally from H$_2$O, CO$_2$ and CH$_4$) are also used
to provide a wavelength calibration accurate to $\simeq$1 km s$^{-1}$.

Unlike other visible and infrared spectrometers, TEXES can provide
an absolute radiometric calibration of line intensities. The method makes
use of observations of ambient and sky loads to
compute the system throughput and gain, as well as the absorptivity of
the atmosphere \citep{lacy}.  For maps of point sources or for 
line sources with extents 
significantly larger than the slit width, this radiometric calibration
does not need to be corrected for slit losses. 

We used maps of $\beta$Gem and $\alpha$Tau to confirm the absolute
calibration scale.  Scans
across $\beta$Gem and $\alpha$Tau (at 12.3 $\mu$m) yielded 
radiometric flux densities of 8.0 and 44.9 $\times$ 10$^{-22}$ 
ergs cm$^{-2}$ s$^{-1}$ Hz$^{-1}$, respectively.  These values agree
to within 5\% with flux densities derived from the N Band magnitudes of 
$\beta$Gem and $\alpha$Tau  
\citep{tokunaga} extrapolated to 12.3 $\mu$m (8.4 and 43.4 $\times$ 10$^{-22}$ 
ergs cm$^{-2}$ s$^{-1}$ Hz$^{-1}$).

Once properly calibrated spectra were available for each position along the 40\arcsec\ 
scans (or, in the case of the S(4) line, each nod pair), we subtracted a
linear baseline from each spectrum.  The individual spatial scans were then 
shifted in position to correct for telescope drift and spatially coincident
spectra (40 scans in S(2) and 80 scans in S(1)) were averaged together.
To smooth the data, we resampled the S(1) data onto the same grid as our S(2) 
observations and smoothed both datasets with a 1.5\arcsec\ Gaussian.  The 
final spatial resolution along the slit for all 3 ground state H$_2$ lines 
is $\sim$2.0\arcsec, while the 
resolution in the direction of our scan is 2.5\arcsec\ for the 0-0~S(1) data 
and 2.0\arcsec\ for the 0-0~S(2) data.

We determined the statistical uncertainty of the integrated intensities 
in the summed spectra at each position by calculating the rms noise
integrated over a stretch of baseline adjacent to the line emission and 
comparable in width to the region
containing most of the line flux.  Table \ref{observations} lists the 
average uncertainty in the integrated intensities for the different lines. 
The highest signal to noise ratio for any individual 2\arcsec\ resolution 
element is 58 for the 0-0~S(1) spectra and 
86 for the 0-0~S(2) spectra.

\section{Results}
\subsection{Maps of H$_2$}
\begin{figure}
\epsscale{1.0}
\plotone{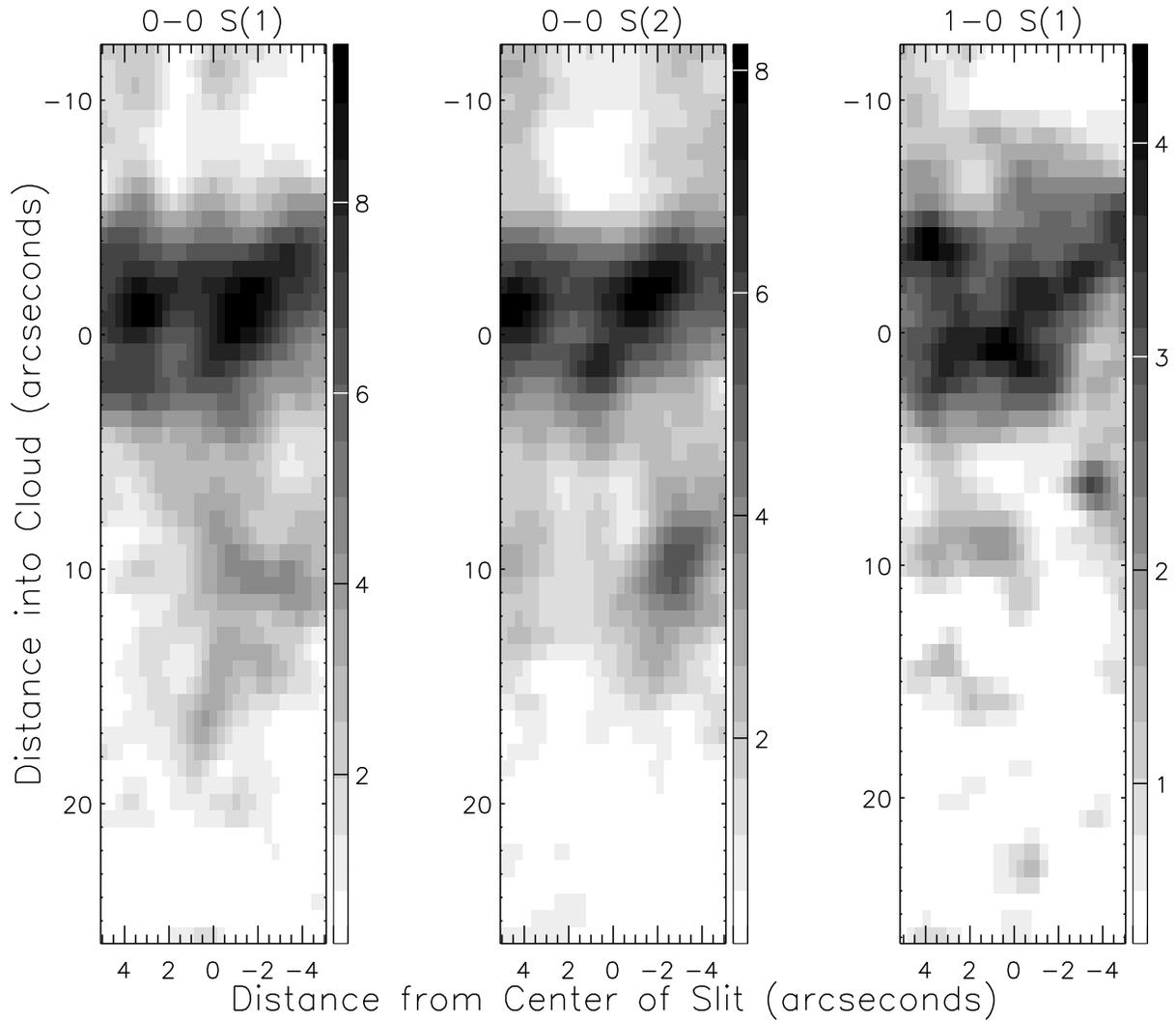}
\caption[1]
%>>>> use \label inside caption to get Fig. number with \ref{}
{\label{maps} Maps of integrated intensities for the S(1) and S(2) pure 
rotational lines taken with TEXES in December 2002, and the v = 
1-0~S(1) line from \citet{vanderwerf}.   
The scans were taken perpendicular to the ionization front at a position angle 
of 45\degree.  The 0,0 position in the maps 
corresponds to R.A. = 5$^h$ 35$^m$ 19.7$^s$, Dec. = -5\degree\ 25\arcmin\ 
28.3\arcsec\ (J2000.0).  Units for labelled values on the greyscale wedges are 
10$^{-4}$ ergs cm$^{-2}$ s$^{-1}$ sr$^{-1}$.  
}
\end{figure}

The first two panels of 
Figure \ref{maps} show the distribution of v=0-0~S(1) and S(2) intensity.
The final panel shows the distribution of intensity in the 2.12 
$\mu$m v = 1-0~S(1) line \citep{vanderwerf} resampled onto the same
grid at the same spatial resolution as our observations of the ground-state
rotational lines. For all three, up is northwest.
The dominant feature in the maps is the bright horizontal (northeast-southwest)
ridge centered at y$\simeq$-2\arcsec.  The overall 
thickness of the bright ridge is $\simeq$8\arcsec\ (0.017 pc).
The brightness of the side of the ridge facing the ionization front
and $\theta^1$C Ori rises very steeply in the ground-state rotational
lines.  The cuts shown in Figure \ref{cuts} indicate that the 0-0~S(1) and S(2)
intensity increases from $<$10\% to 50\% of the peak value in 
2\arcsec--3\arcsec, or barely more than a single resolution element.
In two of the three cuts, the rise toward the peak of the ridge is
much more gradual in the 1-0~S(1) line with emission present at 20-50\% of the 
peak value $\sim$5\arcsec\ in front of the peak.

There is remarkable agreement in the intensity distribution, not only
between the 0-0~S(1) and 0-0~S(2) lines but also between the distributions
in these lines from the vibrational ground state and the distribution of
v=1-0~S(1) emission (Figure \ref{maps}).  The three H$_2$ line maps 
agree to within the 
uncertainties about the location and width of the bright ridge, as well
as the presence and extent of lower-level extended emission behind the ridge. 
Further, there is substructure within the bright ridge that is present in all
three maps.  There are numerous
other small-scale features in the map, with sizes ranging from unresolved
up to $\sim$10\arcsec.  The smaller scale structure in the intensity 
distribution is also evident in the cuts shown in Figure \ref{cuts}.
Along these cuts, displaced by only 3\arcsec\ to the northeast or southwest,
there are substantial differences at adjacent points.

Typical signal-to-noise for the integrated intensities along the bright ridge 
is better than 25 and the peak intensities are 9.6 and 8.2 $\times$ 10$^{-4}$
ergs s$^{-1}$ cm$^{-2}$ sr$^{-1}$ in 0-0~S(1) and S(2) respectively. 
Along the ridge, the line intensities do not drop below 60\% of the peak 
The smaller scale structures behind the ridge have intensities 
that range from 10 to 50\% of the peak. 
The weakest detected emission in both lines is $\sim$0.6 $\times$ 10$^{-4}$ 
ergs s$^{-1}$ cm$^{-2}$ sr$^{-1}$, centered around (+3,+10).  
The signal to noise in the individual 0-0~S(4) spectra is $\sim$20 at the
brightest positions.

\subsection{Spectral Properties}
In order to study the details in the spectral line shapes and to 
compare line shapes from one part of the H$_2$ emission region to another, we 
need to average over larger areas to improve the signal to noise.  We have 
chosen a set of areas that encompass different morphological features 
in the maps of integrated line strength.
We have not placed any of these areas along the leading edge of the 
bright ridge where the steep intensity gradients combined with even our
modest pointing uncertainties make comparisons uncertain.
  Figure \ref{areas} shows the 
outlines of the areas and assigns labels to them.  These labels appear 
then in Figure \ref{spectra} next to the 0-0~S(1) and 0-0~S(2)
(and, for areas A and B, 0-0~S(4)) spectra 
created by integrating over the designated areas. 
With the spectral resolution available with TEXES, we are able to resolve 
the lines.  
Measured FWHM line widths are 5-8 km s$^{-1}$ (Table \ref{velocities}).  
Deconvolving the instrument profile and assuming that the intrinsic line 
shape is Gaussian, we derive physical FWHM line widths of 4-6 km s$^{-1}$ 
(Table \ref{velocities}), in agreement with linewidths expected for optically 
thin, thermal gas at $\sim$1000 K.  The instrumental profile, however, 
may be closer to a Lorenzian, which could imply line widths as low as 
2-4 km s$^{-1}$.  Line widths for the 0-0~S(1) and S(2)
transitions agree to within 1-2 km s$^{-1}$, with neither
line being systematically wider than the other,
while physical linewidths for the 0-0~S(4) spectra are
wider than the 0-0~S(1) and S(2)
linewidths for areas A and B by 1.2~km~s$^{-1}$ and 1.5~km~s$^{-1}$.
In the molecular gas deeper in the cloud, \citet{hogerheijde} and 
\citet{batrla} find linewidths of 1.5 to 5 km s$^{-1}$ (in NH$_3$, 
CS, and isotopes of CO), although deconvolved widths of H$^{13}$CN 
lines toward individual clumps are $<$1 km s$^{-1}$ \citep{lis}.  Closer to 
the ionization front, \citet{wyrowski} measured linewidths of 2 to 
2.5 km s$^{-1}$ 
for C91$\alpha$ emission.  
The radial velocity of the H$_2$ emission (V$_{LSR}$) is 10-11 km s$^{-1}$, 
in agreement with published 
radial velocities for both molecular lines (NH$_3$, CS, and isotopes of CO) 
and carbon recombination lines \citep{hogerheijde,vanderwerf,batrla}.  
Peak velocities of the H$_2$ lines vary by no more than 2 km s$^{-1}$.  

\subsection{Line Ratios and Temperatures}
\begin{figure}
\epsscale{0.4}
\plotone{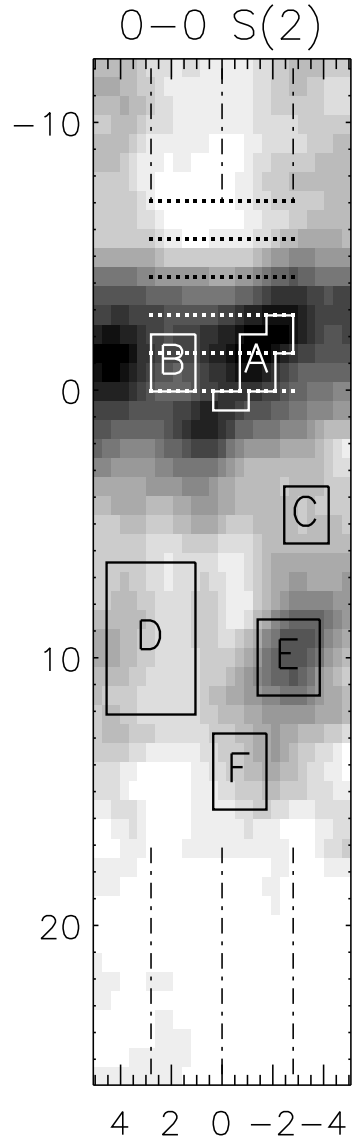}
\caption[1]
%>>>> use \label inside caption to get Fig. number with \ref{}
{\label{areas} Map of integrated intensities for the 0-0~S(2) pure 
rotational lines (identical to the middle panel of Figure \ref{maps}) 
showing 0-0~S(4) slit positions (dotted lines) along with the position of the 
cross cuts in Figure \ref{cuts} (dash-dotted lines) and the 
areas averaged together for further analysis to produce the spectra 
shown in Figure \ref{spectra} and the derived quantities listed in 
Tables \ref{data} and \ref{velocities}.  Distances are in units of arcseconds.
}  

\end{figure}

\begin{figure}
\epsscale{1.0}
\plotone{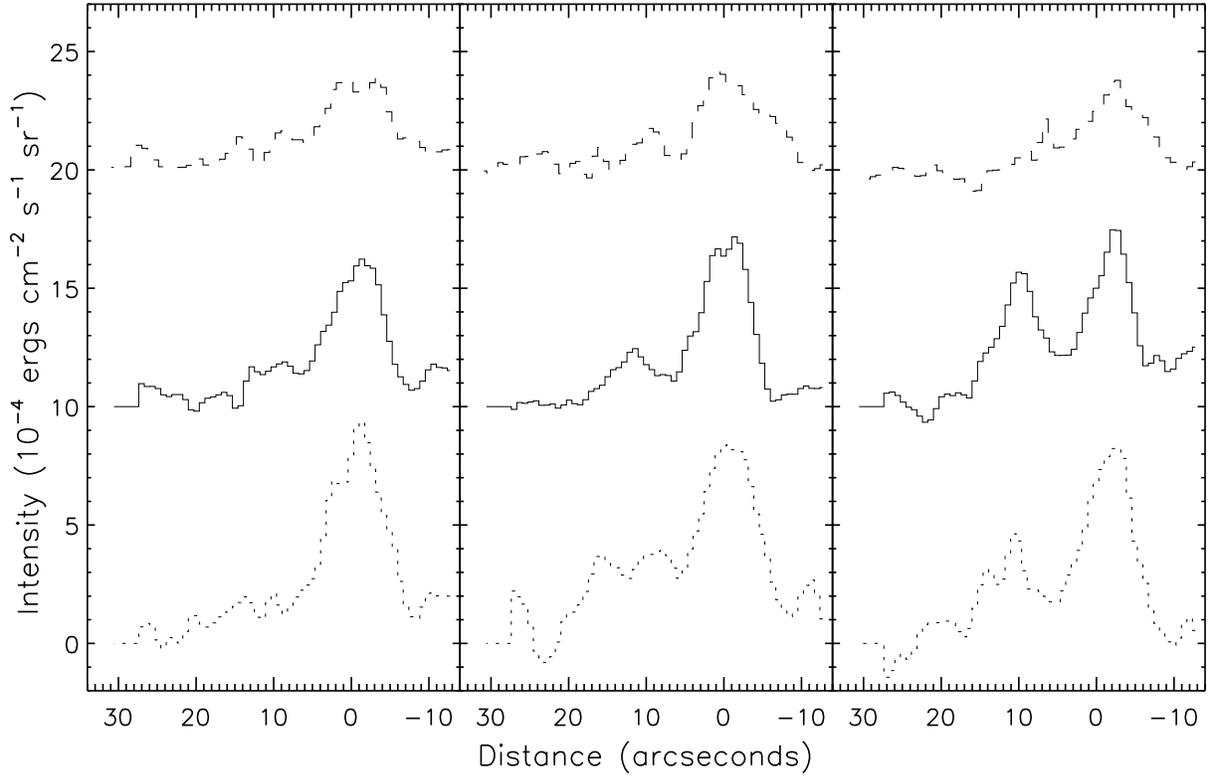}
\caption[1]
%>>>> use \label inside caption to get Fig. number with \ref{}
{\label{cuts} Cuts made through the Orion bar PDR.  
Intensities of the 0-0~S(1) (dotted) 
lines, 0-0~S(2) (solid) lines and 1-0~S(1) (dashed) lines vs. depth into 
the cloud at the center of our slit, 3\arcsec\ southwest along the slit, and 
3\arcsec\ northeast along the slit.  
Distance increases as one moves further from the ionization front (southeast).
The 0-0~S(2) and 1-0~S(1) lines are 
offset by 10 and 20 ($\times$10$^{-4}$ ergs cm$^{-2}$ s$^{-1}$ sr$^{-1}$) 
respectively. The 1-0~S(1) line data are from \citet{vanderwerf}
}  
\end{figure}

\begin{figure}
\epsscale{0.6}
\plotone{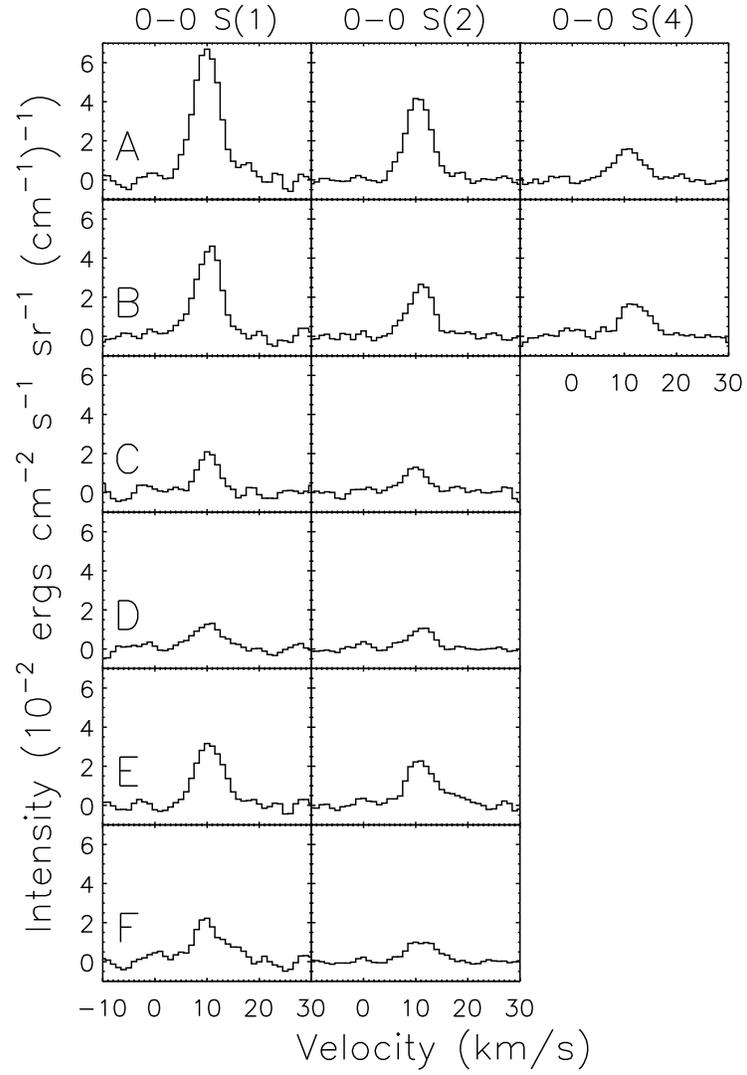}
\caption[1]
%>>>> use \label inside caption to get Fig. number with \ref{}
{\label{spectra} Spectra for summed positions A to F as shown in Figure \ref{areas}.
}  
\end{figure}

Density estimates for the H$_2$ emission zone in the Orion Bar are 
5-25 $\times$ 10$^4$ cm$^{-3}$ derived from C91$\alpha$ observations 
\citep{wyrowski}.  At densities in this range, the upper states of the 
0-0~S(1), S(2), and S(4) transitions should all be thermally populated 
\citep{mandy,bertoldidraine}.  Table \ref{data} lists line intensities 
and excitation temperatures derived from the 0-0~S(1)/0-0~S(2) 
line intensity ratios (assuming that foreground extinction is negligible 
and that the lines are optically thin) for the six areas shown in Figure
\ref{areas} and the excitation temperature derived from the 0-0~S(4)/S(2) line 
intensity ratios for areas A and B.  We use the 
ratio of the 0-0~S(1)/S(2) intensities, together with the measured 0-0~S(1) 
line intensity to derive the column density of warm molecular hydrogen 
toward regions A through F.  We assume that the states are populated 
according to a thermal distribution,
which for T $\gtrsim$ 300 K implies an ortho-to-para ratio of 3.
Even if the averaged regions contain cloud material with temperature gradients,
these derived, single temperature LTE values provide a way of comparing
aggregate properties in the different regions.

Derived excitation temperatures range from roughly 400 to 600 K.  
Along the bright ridge, the temperature varies between 400 and 500 K.  The two
areas in Figure \ref{areas} illustrate this with the bright part of the ridge 
(position A) having T[S(2)/S(1)]= 460 K while the fainter region along 
the ridge (position B) has a slightly lower temperature (430 K).  The derived
column densities at the two ridge positions are very
similar ($\sim 9 \times 10^{20}$ cm$^{-3}$, Table \ref{data}).
T[S(4)/S(2)], however, is cooler in position A
(503 K) than in position B (572 K). There is an apparent trend
(with the exception of position F) toward higher temperatures at greater
depth into the molecular cloud (farther from $\theta^{1}$C).

Our mean temperature for the gas emitting the H$_2$ rotational lines
is similar to that derived by \citet{habart} from ISO observations and 
by \citet{parmar}.
We find only a small overall range in 0-0~S(2)/S(1)
temperatures and do not see the gradient with distance from $\theta^1$C
that was present at modest significance in the \citet{parmar} results.  
\citet{parmar} present spectra integrated along a 
2\arcsec $\times$ 10\arcsec\ slit oriented at the same position angle
as our map but took data along a different cut through the ionization
front.  Their spectra sample the H$_2$ distribution every 5\arcsec\ 
perpendicular to the front.  Our maps indicate that Parmar et al.'s
high temperature point, which lies closest to the ionization front,
is very sensitive to relative pointing errors because of the 
sharp intensity gradients.  If the front position is indeed right at 
the leading edge of the H$_2$ distribution, the second point in
the earlier map may lie behind the brightest part of the H$_2$ ridge. 

Comparing our results to the 1-0~S(1) intensities derived from the
maps of \citet{vanderwerf}, we find values of I(1-0~S(1))/I(0-0~S(1))
of 0.42 and 0.44 at positions A and B, respectively.  Shifting the 1-0~S(1) 
line map by 1.5\arcsec\ produces less than 10\% changes in the 
line intensities at positions A and B.  At position A,
the v=2-1~S(1)/1-0~S(1) intensity ratio is $\sim 0.14$ \citep{vanderwerf}.
At the positions behind the ridge, I(1-0~S(1))/I(0-0~S(1)) varies
from $<$0.07 to 0.65.  Note, however, that the intensity
calibration of the v=2-1~S(1) results is less reliable than the
1-0~S(1) intensities toward positions A and B.  The total area mapped in 
v=2-1~S(1) by \citet{vanderwerf} was only 40\arcsec\ by 40\arcsec\ .  Unlike
the mid-IR spectroscopic mapping results, the on and off line
wavelengths were observed separately in the near-IR so that zero
point offsets could occur that would have a stronger effect on the
line calibration at points where the lines are weak.  If, as appears 
to be the case, the intensities were set to zero at the edge of the field 
mapped in v=2-1~S(1), then any flux on scales greater than 40\arcsec\ 
would be absent.  This problem is less acute for the v=1-0~S(1) map both 
because the intensity distribution peaks more sharply and because the 
\citet{vanderwerf} map of this line covers a considerably larger area.  
\citet{usuda_etal_1996} also used Fabry-Perot imaging to determine 
the 2-1~S(1)/1-0~S(1) ratio.
Although their spatial resolution of 8\arcsec\ may fail to resolve some of
the structure, their larger map area results in a more reliable flux ratio.
The \citet{usuda_etal_1996} v=2-1~S(1) map implies that the edges of the 
\citet{vanderwerf} field lie at about the 50\% level of the line intensity 
distribution.  Given the observing and data analysis techniques, the two 
v=2-1~S(1) results are consistent with one another, but the 
\citet{usuda_etal_1996} value for the 2-1~S(1)/1-0~S(1) ratio is more reliable.
We therefore use their value, $0.25$, for the 2-1~S(1)/1-0~S(1) intensity
ratio.
We discuss the observed H$_2$ line intensity ratios in the context of 
a realistic model of the temperature structure in \S 4.

The formal uncertainties in the LTE temperatures, based on the signal to 
noise of our ground vibrational state H$_2$ spectra are small.  
The maximum uncertainty in temperature was 40 K for the temperature derived 
from the 0-0~S(1)/0-0~S(2) line ratio in region D (Figure \ref{areas}).
Regions with brighter lines (areas A, B, \& E) had much lower uncertainties
(4-10 K).  
Registration and positional errors had a modest effect on the derived 
average temperatures.  We shifted the 
0-0~S(1) map by +0.7\arcsec\ and -0.7\arcsec\ (greater than our
quoted position uncertainty of 0.5\arcsec).  The maximum change in temperature
due to these shifts was 80 K for temperatures derived from
the 0-0~S(1)/0-0~S(2) line ratio for the
summed spectra in region C.  Typical changes in derived temperature from the 
0-0~S(1)/0-0~S(2) line ratio due to a comparable position shift were 25 K. 
At the leading edge of the bright ridge, however, temperatures could change
from $\sim$500 to $\sim$900 K with relative position shifts of $\sim$3\arcsec.

We do not correct for extinction in our derivation of temperature, as
the effects of reddening are expected to be quite low.
In Orion OMC-1, \citet{rosenthal} find values of 
$\rm A_{\lambda}/\rm A_{\rm K}$ of 0.525,
0.527, and 0.441 for the 0-0~S(1), S(2), and S(4) lines respectively.  
Using these values for A$_{\lambda}$, and assuming 
A$_{\rm K}$/A$_{\rm V}$=0.112 
\citep{rieke}, we find that an error in T of 50 K corresponds to 
A$_{\rm V}\sim$20, for temperatures derived from the 0-0~S(1)/S(2) and 
0-0~S(4)/S(2) line intensity ratios.

A single temperature cannot describe our observed line intensities (to within 
the uncertainties) for the pure rotational lines.  The deviation from a single 
temperature fit becomes more severe as we include vibrationally 
excited lines.  Predictions of the 1-0~S(1) 
line intensity based on the temperatures and column densities derived from 
the rotational lines for areas A \& B underestimate the observed 1-0~S(1) line 
intensity by factors of 4 to 75.  The enhanced 1-0~S(1) line intensities could 
be caused by a temperature gradient in the region of H$_2$ emission or 
by fluorescent excitation of the vibrational lines.  Clearly, more complex 
analysis of line ratios is necessary.  We discuss the inputs and results of 
our PDR model analysis in the next section.

\section{Discussion}
 
\subsection{Inputs to a PDR Model for the Orion Bar}

The Orion bar is a particularly useful test site for PDR models
because of its closeness \citep[at 450 pc 2\arcsec=1.4$\times10^{16}$ cm,][]
{hoogerwerf}, its high gas density and strong incident UV field (leading
to bright line and continuum emission), and its distinctive
geometry.  These favorable properties have inspired a significant
number of observational studies over the past two decades which
can provide valuable inputs into any model of the region.
The key region-specific parameters in modeling photodissociation regions
are the strength of the incident UV field and the density of the
neutral gas as a function of distance from the front.
In comparing the models to observations, it
is also important to understand the source geometry, in particular
the tilt of the source with respect to the line of sight.

The usual practice in modeling photodissociation regions has been to
assume that the dust-related parameters in the models: dust
opacity, photoelectric heating rate, and H$_2$ formation rate, are
fixed  and to search through a family of models in an attempt to
find the best match when the incident UV field and the 
density (often taken to be uniform) of the PDR
are varied \citep{burton90}.  Because the UV
field incident on the Orion Bar and the pressure at the ionized
boundary of the PDR are well constrained by observations, we can
reverse the usual procedure, take the UV field and pressure as
givens, and use Orion as a testbed for the dust-related parameters 
going into models of high density PDRs with high incident UV fields.  
In a companion paper
(Draine et al. 2005, Paper II), 
we discuss this model study in detail. In the
current work, we discuss the derivation of the initial conditions
from observations and compare the results of our observational
study to the best-fit PDR model derived in Paper II.

The dominant source of UV photons for the Orion Bar PDR
is the O6 star $\theta^1$C Ori.  At the position of our
H$_2$ map, $\theta^1$C Ori lies 120\arcsec\ from the ionization
front where the direction of the incident radiation is only 20 degrees
from the normal to the front projected onto the plane of the sky.
It is convenient to measure the intensity of the radiation incident
at the ionization front by $\chi$, the ratio of specific energy
density at 1000~\AA\ to the value $u_\lambda$= 4$\times 10^{-17}$ ergs
cm$^{-3}$ \AA$^{-1}$ estimated by \citet{habing} for the mean
interstellar radiation field.  Based both on simple geometric
dilution and on the strength of the far-IR radiation emitted by
the warm dust in the Orion bar after being heated by the UV
and visible radiation from $\theta^1$C Ori, the far-UV flux
incident on the PDR is $\sim$3$\times$10$^4$ times the mean
interstellar radiation field \citep{herrmann}.  
\citet{marconi} used observations of the OI
1.317 $\mu$m line to infer the incident UV intensity at 1040 $\AA$.
If the PDR is inclined with cos$\theta$= 0.1, where $\theta$ is the angle 
between the line of sight and normal to the PDR, the Marconi results
imply $\chi$= 2.9$\times$10$^4$.  For all of the models explored
in Paper II, we take $\chi$= 3$\times$10$^4$ at the ionization front.

$\theta^1$ Ori C, A, and E combined have a 1--10~keV luminosity
$L_X(1-10~{\rm keV})=2.4\times10^{32}{\rm erg~s}^{-1}$ 
\citep{Schulz_etal_2003}.
At a distance of $\sim8\times10^{17}$~cm distance, 1--10~keV X-rays from
the Trapezium stars will contribute an ionization rate
$\sim10^{-16}$~s$^{-1}$ at the Bar.  Lower energy X-rays from
the Trapezium, and X-rays from young stars
that are less luminous but are closer to the Bar will contribute
additional ionization.
In addition, the cosmic ray ionization rate may be enhanced in this
region by nonthermal particle acceleration in stellar wind shocks.
We adopt a nonthermal ionization rate 
$\zeta_{CR}\approx 1\times10^{-15}$~s$^{-1}$ for gas in the Orion
Bar PDR, but we stress that our results do not depend sensitively on this rate.
We note that \citet{McCall_etal_2003} inferred an ionization rate
$\sim1.2\times10^{-15}$~s$^{-1}$ in the molecular gas toward $\zeta$~Per.

Based on the emission measure derived from radio continuum observations
\citep{felli}, and $n_e$ determined from [S~II]$I(6716)/I(6731)$ 
\citep{Pogge etal 1992}, the thermal pressure at the ionization front
$nT\simeq 6\times10^7$~cm$^{-3}$~K.  The gas has been accelerated
away from the PDR, however, so the pressure in the PDR should be
somewhat larger. For the PDR models of the Orion Bar, we have taken
the total pressure to be uniform through the PDR at 
$P/k$ = 8$\times$10$^7$~cm$^{-3}$~K.  In most
previously published PDR models dealing with H$_2$ excitation
\citep{black,sternberg89,burton90}, the assumption has been that
turbulent or magnetic pressure dominate 
the gas pressure throughout the region and that an
assumption of constant density is therefore reasonable.  
The Draine et al. (2005)
models explicitly calculate density as a function of depth into the PDR
assuming a constant pressure that includes a 
non-thermal contribution\footnote{%
   The non-thermal pressure is taken to be $p_{\rm nt}=\rho v_{\rm nt}^2$,
   with $v_{\rm nt}=1 {~\rm km~s}^{-1}$.}
that is
fairly small in the outer parts of the region. 
In the outer parts of the PDR, constant densities derived from older
PDR models and various observations are generally consistent with the
pressure value used in our models, albeit with a large spread.
The densities derived by
\citet{wyrowski} from the C91$\alpha$ results (5--25 $\times$~10$^4$~cm$^{-3}$)
are consistent with the assumed pressure if T$\sim$1000 K in the carbon
line formation region. 
Non-LTE excitation modeling of millimeter and submillimeter molecular
line ratios is not dependent on PDR model results and yields
densities ranging from a few 10$^5$ cm$^{-3}$ to
a few 10$^6$ cm$^{-3}$ \citep{burton90,tauber95,hogerheijde,youngowl} 
in the gas farther behind the ionization
front where molecular line ratios and brightness temperatures imply
$T\approx120$~K.  The higher densities are in good
agreement with a virial analysis of the brightest HCN clumps
in the molecular ridge \citep{lis}.

The pressure assumed for our models is also consistent with densities
derived by a second line of argument, based on geometry and the
chemical stratification shown in older constant density models of
PDRs.  There is a clear
stratification of emission zones, manifested by a shift in
the observed location of the peak emission, as one goes farther
from $\theta ^1$C into the PDR, albeit with some overlap
of what should be, from a theoretical point of view, distinct
regions within the PDR (Figure \ref{context}).  Along a line perpendicular
to the bright ridge in Figure \ref{maps}, the ionization front lies at 
$y\approx-17$\arcsec\ \citep{felli}. Within the neutral gas,
there are successive emission zones for the FeII 1.64 $\mu$m line
\citep[$y\approx-16$\arcsec]{marconi}, the 3.3 $\mu$m PAH feature 
\citep[$y\approx-12$\arcsec]{bregman}, 
the H$_2$ rotational and ro-vibrational transitions (this paper, 
$y\approx-2$\arcsec), submillimeter continuum emission 
\citep[$y\approx+5$\arcsec]{lis98}, 
and various millimeter and submillimeter lines of HCO$^+$, CO, and HCN 
\citep[$y\approx\!+8$\arcsec]{tauber,hogerheijde,youngowl,lis}.
The peak of the H$_2$ rotational line emission in Figure \ref{maps} is at
$y\approx-2$\arcsec, 
or $\sim$15\arcsec\ ($\sim$9$\times10^{16}$cm) from the ionization front.
By comparing the linear displacements of these peaks to the
column density peaks in plane-parallel PDR models with
appropriately chosen incident UV fields, \citet{tielens,tauber}
and \citet{simon} estimate the density for
a homogeneous medium to be 5$\times$10$^4$ to 3$\times 10^5$
cm$^{-3}$, consistent with the range of
densities derived from physical measurements and with an
appropriate density range for our constant pressure models.

Many authors have suggested that propagation of far-UV
radiation through a clumpy medium offers an explanation
for the range in derived densities, particularly in the molecular part of 
PDR's.  
In such a picture, the attenuation scale length in the low density medium
or the distance to reach a clump area filling factor of
unity set the size scale for the PDR \citep{stutzki}.  
There is, however, no unambiguous 
evidence of a clumpy structure in the region of the Orion Bar where 
H$_2$ emits.
The fairly straight leading edge of the bright H$_2$ ridge
and the absence of strong variations in either column
density or in the ratio of 1-0~S(1) line to 0-0~S(1) line
intensity along the ridge argue that the H$_2$ emission
arises either from a uniform medium, from a low density PDR component, 
or from an ensemble of clumps with a line of sight filling factor
significantly larger than one.  Clumps with densities of $\sim$6 $\times$ 
10$^6$ cm$^{-3}$ have been observed deep into the molecular gas behind the PDR
via HCN, HCO and their 
isotopomers \citep{lis,youngowl}.  However, for the neutral gas closest to the 
ionization front, \citet{marconi} use
the relative strengths of near-IR Fe~II lines to exclude
the presence of clumps with densities $>$10$^6$~cm$^{-3}$.  
In our models (Paper II), we therefore assume a uniform medium.

Apart from the temperature within the H$_2$ bright ridge
of $\sim$450 K, there are several other temperature measurements
that can serve as constraints on the thermal balance through
the PDR.  If the carbon recombination lines arising from
the part of the PDR closest to the ionization front are
purely thermally broadened, the temperatures in that layer
are 1000-1600 K \citep{wyrowski}.  CO 6-5 brightness
temperatures $\sim$10\arcsec\ in front of the molecular peak
($\sim$5\arcsec\ behind the H$_2$ ridge) imply a kinetic temperature
of 120-180K \citep{lis98}. Farther into the cloud,
\citet{batrla} use NH$_3$ line ratios to derive
a temperature of 120 K for what they argue are the surfaces
of the dense clumps in the molecular ridge.

In comparing the PDR model intensities to observed intensities,
we must tilt the models correctly with respect to the line of
sight and account for radiative transfer through the inclined
PDR slabs.  There are two lines of evidence that argue that the Orion
bar is highly inclined from the plane of the sky.  Molecular
line observers, who see low-level emission both in front
of and behind the bar, conclude that the roughly factor
of 10 enhancement in column density seen for optically
thin lines from the bar implies an inclination of only a
few degrees from the line of sight \citep{tauber,hogerheijde}.  
The steepness of the dropoff in radio continuum at the ionization front
and the steep rise in the rotational H$_2$ line emission
on the leading edge of the ridge form the second argument
for the almost edge-on orientation of the PDR \citep[this paper]{felli}.
\citet{walmsley} have estimated that the Orion bar is
approximately plane-parallel and that we view it from a direction
with 1/cos$\theta$ = 10 where $\theta$ is the angle between the
line-of-sight and the normal to the PDR.  In calculating models of
the emergent line intensities, we adopt this as a plausible estimate
for the enhancement of the surface brightness of optically thin
lines relative to the face-on surface brightness.  Accordingly,
in all of the models in Paper II, Draine et al. attempt to reproduce the
line intensities observed at positions A and B with a plane-parallel
PDR viewed from an angle such that 1/cos$\theta$ = 10.  
The model line intensities include attenuation by dust 
within the PDR.\footnote{%
    The dust attenuation cross section is taken to be
    $\sigma_\lambda = (A_\lambda/A_{1000{\rm\AA}})\sigma_{1000}$, where
    $A_\lambda$ is the extinction at wavelength $\lambda$ for an
    $R_V=5.5$ extinction law.
    }
For 1/cos$\theta$ = 10, 
internal extinction significantly attenuates the emission at the shorter 
wavelengths.  For example, in the PDR model discussed below
the 1-0~S(1) line is attenuated by a factor of 0.26, the equivalent of A$_{\rm K}$ = 1.4\ .  The lower
surface brightness seen at positions C-F may be due to viewing
with a different inclination angle \citep{hogerheijde}, or possibly
this emission arises from a region physically separate from the location
of the edge-on Bar.  

\subsection{Results for the Best PDR Model}

We present here a comparison of our new observational results for
the Orion Bar and a theoretical model for this high-excitation PDR.
This model may be relevant not only for a single dense PDR illuminated
by O and B stars
in galactic star forming regions but also for studies of physical
conditions over large areas in the inner regions of starburst
galaxies.  The 0-0~S(2)/0-0~S(1) line ratio for the nucleus of
NGC 253 \citep{devost} is the same as observed in the Orion Bar
and the intensity averaged over an 800 pc $\times$ 700 pc region
of NGC 253 is fully 50\% of that observed at peak A in the Orion
Bar. If the emission in NGC 253 originates from PDRs,
these must be both very intense and have a high surface filling
factor.

In paper II, Draine et al.\ present a 
grid of models for the Orion Bar PDR near position A.  
Table \ref{modelin} gives the values of the pressure ($P$), 
radiation intensity ($\chi$),
rate of ionization of H by cosmic rays or X-rays ($\zeta_{CR}$),
and abundances of the coolants C, O, Si, and Fe used in the models.
Gas-phase abundances for C, O, Si, and Fe are taken from 
\citet{Jenkins_2004}
for gas with ``depletion factor'' $F=1$, corresponding to approximately the
level of depletion seen in the diffuse molecular cloud toward $\zeta$~Oph. 
The vibrational line emission from the models is sensitive to the
rate coefficients for vibrational deexcitation of H$_2$, particularly
by collisions with H atoms. \citet{usuda_etal_1996} found that in the 
Orion Bar, the 2-1~S(1)/1-0~S(1) intensity ratios were anticorrelated 
with the 1-0~S(1) line intensities, and were usually lower 
than the 2-1~S(1)/1-0~S(1) intensity ratio ($\sim$0.6) expected for 
pure UV-fluorescence.  
Our values adopted for these rates are
discussed by Draine \& Bertoldi (2005, in preparation).
Table \ref{collision} compares different estimates for the $T=1000$~K rate
coefficients, $k_{\rm vdexc.}(v,J)$, for vibration deexcitation by H atom 
collisions of the $(v,J)=(1,3)$ and $(2,3)$ levels of H$_2$ 
(the levels responsible for 1--0S(1) and 2--1S(1) line emission).  
Our adopted rates are an order of magnitude smaller than the
vibrational deexcitation rates adopted by \citet{sternberg89}, but
exceed the rate coefficients calculated by \citet{mandy},
by factors of 8 and 2, respectively.  Our rates are a factor of
150 times larger than the rates recommended by \citet{lebourlot}.

The model grid explores variations in the dust ultraviolet attenuation
cross section, the H$_2$ formation rate, and the photoelectric heating rate.  
Table \ref{modelout} compares 
a model from this grid (Model 1) to the observations at position A.
The model is within $\sim$10\% of the 0-0~S(1), 0-0~S(2), 0-0~S(4),
and 1-0~S(1) intensities for the average of positions A and B.
The model 0-0~S(1) intensity is $\sim$2\% below 0-0~S(1) at position A, and
$\sim$9\% above the value at B.  For 0-0~S(2), the 
model is $\sim$18\% below A and $\sim$1\% below B.
For 0-0~S(4), the model is $\sim$7\% above A, and $\sim$14\% below B.
For 1-0~S(1), the model is $\sim$6\% above A, and $\sim$12\% above B.

The 2-1~S(1)/1-0~S(1) intensity ratio for the model is 0.23.
There is some uncertainty concerning the observed line ratio.
As discussed in \S 3.3, we use the results of 
\citet{usuda_etal_1996}, 2-1~S(1)/1-0~S(1) = 0.25 at position A, 
which we take as the best observational determination.

\begin{figure}
\epsscale{0.6}
\plotone{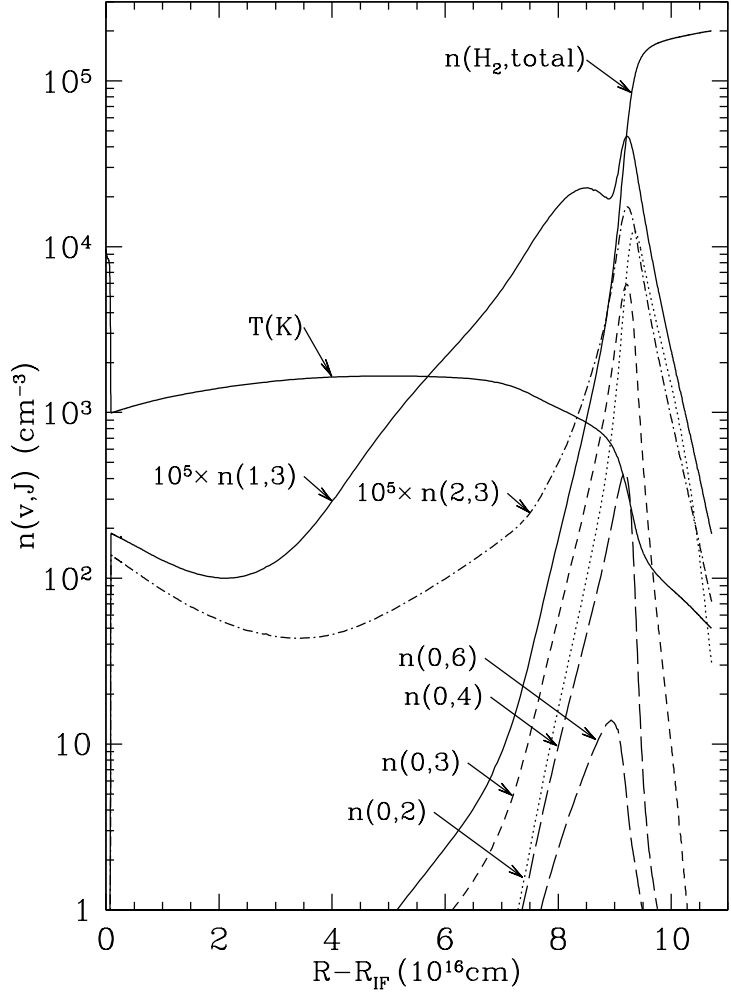}
\caption[1]
%>>>> use \label inside caption to get Fig. number with \ref{}
{\label{model} Distribution and excitation of H$_2$ for Model 1 of Draine
et al. (2005). The plot shows the run of temperature $T$
and of the volume density $n(v,J)$ of H$_2$ molecules in selected $(v,J)$
states, as well as the total H$_2$ density as a function of
distance from the ionization front. The populations in the upper states of
the 0-0~S(1), 0-0~S(2), and 1-0~S(1) transitions all peak just as the
H$_2$ abundance becomes significant and the temperature begins to drop
sharply.  The density $n(2,3)$ of the level responsible for 2--1~S(1) emission
peaks at $\sim9.2\times10^{16}\,$cm, because the falling gas temperature
and declining H abundance reduce the rate of collisional deexcitation
of UV-pumped vibrationally-excited levels; this peak accounts for
most of the column density $N[{\rm H}_2(2,3)]$; the corresponding peak in
$n(1,3)$ accounts for about half of the column density $N[{\rm H}_2(1,3)]$,
with the remainder largely due to thermal excitation in the portion of
the PDR that is primarily atomic.
}  
\end{figure}

Figure \ref{model} shows the temperature profile for Model 1.  The ionization 
front is defined to be the point where $n({\rm H}^+)=n({\rm H}^0)$; 
at this point, the gas temperature is $\sim$9000~K, but the temperature 
drops rapidly with distance from the ionization front, 
as heating due to photoionization of H declines and the fractional 
ionization drops.  
Model 1 successfully reproduces the observed $\sim9\times10^{16}$~cm 
separation between the ionization
front and the peak of the H$_2$ line emission (Figure \ref{model}).  In fact,
the figure shows that the model even has an extended (2-3\arcsec) tail
on the ionization front side of the 1-0~S(1) peak, consistent
with the observed cuts shown in Figure \ref{cuts}.  
Most of the 1-0~S(1) emission in this model arises from collisional
excitation of (1,3), the $v=1,J=3$ state.  
For example, at $R-R_{\rm IF}=6.0\times10^{16}$~cm,
$n(1,3)/n(0,3)\sim0.023$, which is essentially the thermal ratio
($e^{-5936{\,\rm K}/T}$) at the local temperature $T=1600\,$K.
The density is not high enough to fully thermalize the
vibrational levels -- ultraviolet pumping contributes in part to
the population of
(1,3), and accounts for most of the population of (2,3).
The rise in $n(1,3)$ to a local maximum at $8.5\times10^{16}$ cm reflects
competition between increasing $n({\rm H}_2)$ and declining $T$.
The second maximum at $9.2\times10^{16}\,$~cm is due to UV-pumping: the
decline in $T$ and the drop in $n({\rm H})/n_{\rm H}$ lead to a drop in
the rates for collisional deexcitation of the vibrationally-excited states,
so that the $v=1$ levels are no longer collisionally deexcited, 
but going deeper into
the cloud, the UV pumping rates drop and therefore so does $n(1,3)$.
Note this maximum of $n(1,3)$ coincides with the maximum in $n(2,3)$.
 
The rotational levels of $v=0$ 
are thermalized and arise in the zone where the temperature gradient is quite 
steep. The lower $J$ levels [e.g., (0,2)]
peak farther from the ionization front 
than the higher $J$ levels [e.g., (0,6)].

\subsection{Parameters for the Best PDR Model}

The best-fit model from Paper II (Model 1) 
uses a rate coefficient
for formation of H$_2$ on dust grains
\footnote{%
    With the usual definition:
    $(dn({\rm H}_2)/dt)_{\rm form}=R_{{\rm H}_2}n_{\rm H}n({\rm H})$.
    }
with a value $R_{{\rm H}_2}=3.8\times10^{-17}$cm$^3~$s$^{-1}$
at $T=1000$~K -- 
similar to the value $3\times10^{-17}{\,\rm cm^3\,s^{-1}}$
found by \citet{habart} for the Orion Bar PDR.
However, Model 1
implies significant deviations from
the standard values adopted for other dust-related parameters in the 
Orion Bar PDR.

In order to achieve the
agreement in the separation of the ionization front and H$_2$ peak,
Model 1 adopts a dust attenuation cross section
at 1000\AA\  $\sigma_{1000}=0.48\times10^{-21}$cm$^2$ -- if a significantly
higher value of $\sigma_{1000}$ is used, the 
increased FUV attenuation brings the H$_2$ peak too close
to the ionization front.  The adopted $\sigma_{1000}$ is significantly
smaller than the $\lambda=1000$\AA\ 
extinction cross section $\sim2.3\times10^{-21}$cm$^2$ 
inferred from the \citet{fitzpatrick99} parametrization of the
interstellar reddening law for sightlines with
$R_V\equiv A_V/E(B-V)\approx 5.5$, if we take
$N_{\rm H}/E(B-V)\approx 5.8\times10^{21}{\,\rm cm}^{-2}$ from \citet{bohlin}.
At 1000\AA\ the dust albedo is estimated to be $\lesssim0.4$ 
\citep{draine2003a,gordon2004}, implying an attenuation cross section 
$\gtrsim0.6\times2.3\times10^{-21} {\rm cm}^2 
= 1.4\times10^{-21}{\rm cm}^2$ -- 3 times
larger than the value adopted for Model 1.
The dust now in the Orion PDR might have
undergone extensive coagulation during the long time it spent in
cold, dense molecular gas prior to the arrival of the photodissociation
front.
Such coagulation would lower the far-ultraviolet
scattering and absorption per H nucleon.

As described above, 
the PDR model corrects for internal absorption in the PDR assuming
a $R_V=5.5$ reddening law, and therefore the K band attenuation
coefficient has been scaled down by the same factor of $\sim3$
as the UV extinction.  This would be appropriate if the reduced extinction
were due to an overall deficiency of dust grains, but
it would not be correct if the low UV extinction were due to dust
coagulation, as coagulation of small grains 
would not decrease the K band extinction unless
the coagulation resulted in grains larger than $\sim1\mu$m.
As noted above, even the reduced extinction assumed in the model has
attenuated the 1-0~S(1) line intensity by a factor 0.26 because
we assume that we are observing the PDR from a direction with
$1/\cos\theta=10$ -- if the K band
attenuation coefficient were significantly larger than the (reduced) value
in the model, it would be very difficult to reproduce the observed
H$_2$ line intensities.
The degree to which the observed line intensities have
been affected by extinction in the Orion Bar is an important question;
additional observational studies of the reddening using H$_2$ emission lines
would be of great value.

In order to lower the 2-1~S(1)/1-0~S(1) line ratio from the pure
fluorescence value $\sim$0.6 to the observed value
$\sim$0.25, the atomic zone of the PDR must have a gas temperature
$T\gtrsim 1000$K -- this is required so that 
(1) the rate coefficients for collisional
deexcitation are large enough to suppress 2-1~S(1) emission by
collisionally deexciting H$_2$ in the $v=2,J=3$ state fast enough to
compete with spontaneous decay,
and (2) to collisionally excite 1-0~S(1) emission.
Although the models explicitly include heating from collisional deexcitation
of vibrationally-excited H$_2$, the dominant heating process is
photoelectric heating from dust.
With the cooling processes that are present, the only way to
produce the required high temperature is for the heating rate to
be substantially larger than the photoelectric heating rate predicted
by existing models of photoelectric emission from dust 
\citep{bakes,weingartner+draine_2001}.
Draine et al.\ (2005) provide this additional heating
by means of an ad-hoc increase in the photoelectric heating rate
in the atomic portion of the PDR by a factor $\sim3$ relative to the estimate
of \citet[][hereafter WD01]{weingartner+draine_2001} for $R_V=5.5$ dust.

Above it has been argued that the separation of the peak of the H$_2$
emission from the PDR requires a reduction in the FUV absorption by the
dust; since photoelectric heating cannot occur without absorption of
UV photons, one might have expected a corresponding 
reduction in the dust photoelectric heating rate,
whereas Model I posits an increased heating rate.
The increased photoelectric heating rate may be regarded as a proxy for
some other heating process that may be present, or perhaps it is indicative of
overestimation of the fine structure cooling (dominated by [OI]63$\mu$m
and [CII]158$\mu$m emission).
In any event, it indicates that there is a substantial error in our
account of the heating and cooling in the atomic zone of the PDR.

However, although high temperatures, and therefore an enhanced
photoelectric heating rate or its effective equivalient, 
are required to suppress 2-1~S(1) emission in the
region where UV pumping of H$_2$ is taking place, this enhanced
photoelectric heating rate cannot be present in the
regions that are predominantly molecular -- otherwise there would
be too much emission in 0-0~S(4), 0-0~S(2), and 0-0~S(1).
Draine et al.\ (2005) therefore adopt an ad-hoc photoelectric
heating rate that is reduced to $\sim0.4$ of the WD01 heating rate where
$2n({\rm H}_2)/n_{\rm H}=0.5$, and $\sim$0.1 of the WD01 heating rate where
$2n({\rm H}_2)/n_{\rm H}=0.9$.
Such variation in the grain photoelectric heating properties
could perhaps come about if the grains from the cold dense molecular cloud 
enter the PDR in some state (perhaps coated or clumped) yielding
a low photoelectric heating rate.
As these grains enter the PDR and are exposed to both the $\lambda<1100$\AA\
radiation radiation that dissociates H$_2$ 
and atomic H, perhaps the grain properties are altered
(e.g., dispersal of clumps, or photolysis of coatings) so as to
increase the photoelectric yields.

%The deconvolved linewidths for our H$_2$ ground state rotational lines are 
%4-6 km s$^{-1}$, in agreement with linewidths expected for optically thin, 
%thermal gas at $\sim$1000 K. 
%However, the line ratios dictate lower gas temperatures in the region
%accounting for the observed rotational lines.
%In our model, $n(0,6)$ peaks at a point where $T=620\,$K.
%Thermal broadening at this temperature gives FWHM = 3.8 km s$^{-1}$,
%significantly below the observed FWHM = 5.7 km s$^{-1}$ for 0-0~S(4).
%If the observed FWHM is due to thermal plus turbulent velocity dispersions
%adding in quadrature, the implied turbulent FWHM is 4.2 km s$^{-1}$,
%surprisingly large.  Similarly, $n(0,4)$ peaks where $T=240\,$K, for a 
%thermal FWHM = 2.6 km s$^{-1}$, again significantly 
%smaller than than the observed FWHM = 4.5 km s$^{-1}$ for 0-0~S(2).
%If the observed FWHM of 0-0~S(2) 
%is due to thermal plus turbulent velocity dispersions
%adding in quadrature, the implied turbulent FWHM is 3.7 km s$^{-1}$.
%The implied velocity dispersion of the gas -- whether due to turbulence
%or unresolved coherent velocity gradients in the flow -- 
%is surprisingly large, at a level that must be dynamically important.  
%If the motions are due to turbulence or
%waves, perhaps there is heating due to dissipation of this kinetic
%energy that could account for the unexpectedly high heating rate in the
%atomic zone, which in the model is provided by an ad-hoc increase in the
%photoelectric heating rate.

There are no velocity-resolved observations of 
1-0~S(1) toward the Orion Bar.  In the PDR associated with the reflection 
nebula NGC 7023, this line has a width of 3.4 km s$^{-1}$ \citep{lemaire}, 
consistent with our measurements of the ground state lines in the Orion Bar. 
The narrow
linewidths of our pure rotational H$_2$ lines indicate that the gas, 
if shocked, must be shocked at a very low velocity.  Observed 
linewidths in regions with even moderate (v$_s$ $\sim$ 20 km s$^{-1}$) shock 
velocites are greater than $\sim$ 30 km s$^{-1}$ \citep{parmarshock, tedds}.
Shock models of \citet{draineshock} 
and \citet{kaufman} predict H$_2$ v=1-0~S(1) intensities greater than 
 $10^{-4}$ ergs s$^{-1}$ cm$^{-2}$ sr$^{-1}$ in face on PDRs 
from shocks with velocities greater than 20 km s$^{-1}$.
\citet{tielens} calculate the heating input by a shock in the Orion Bar and 
find that heating of the gas by the FUV field (65 erg cm$^{-2}$ s$^{-1}$) 
exceeds shock heating unless the shock velocity is greater than 10 km s$^{-1}$.
Given the morphological similarities of the 1-0~S(1) and 0-0~S(1) emission,
it is also unlikely that shock excitation could appreciably contribute to 
the observed 1-0~S(1) intensity.  

In addition to explaining the PDR structure and H$_2$ line intensities,
future modeling
must address the high temperatures in the CO/HCO$^+$/NH$_3$ zone of the
Orion Bar.  The models in Draine et al. (2005, in prep) 
do not calculate the thermal balance realistically that far into the PDR.  
Figure \ref{model} shows, however, that the model temperature
has already dropped to 50~K, far below the temperatures of 100-120 K
derived from observations of the molecular zone, even before that zone 
is reached.  Clearly, other heating mechanisms must be in play within
that zone as well.  We emphasize that our models of the Orion Bar are 
optimized for the region in the PDR where H$_2$ emits, and do not apply 
to regions deeper into the molecular cloud.

\section{Summary}
We obtained high resolution (R = 75,000 to 100,000) spectral maps 
of H$_2$  v = 0-0~S(1) and S(2) covering a 12\arcsec\ by 40\arcsec\ 
region in the Orion Bar PDR.  Linewidths for the spectra in our maps are 
4-6 km s$^{-1}$ with V$_{LSR}$ ranging from 10.2 to 11.5 km s$^{-1}$.  
Comparison of our maps with v = 1-0~S(1) observations \citep{vanderwerf} 
reveals exceptional similiarity in the line intensity distributions.

To model our line intensities (detailed in Draine et al. 2005, in prep), 
we use 
estimates of the FUV field, pressure and inclination angle from the literature 
(\S 4.1), and allow flexibility in dust-related parameters 
(dust opacity, photoelectric heating rate and H$_2$ formation rate).  
The best-fit model matches the distance between the 
H$_2$ line emission and the ionization front,
the observed intensities of the H$_2$ $v=1-0$~S(1) and $v=2-1$~S(1) lines,
and the intensities of the ground state rotational lines.

In order to reproduce the observed separation between the ionization front
and the H$_2$ emission peak, the model requires a reduction in the FUV
attenuation cross section, by a factor of $\sim$3 relative to
a priori estimates.  This model also requires 
an enhanced heating rate in the 
atomic region of the PDR, corresponding to a factor $\sim3$ increase
in the photoelectric heating rate
(or a corresponding reduction in the radiative cooling)
in order to maintain $T\approx 1500$K in the atomic zone.  Though the 
uniqueness of our solution has not been tested, it is apparent that the 
standard dust-related parameters used in PDR models do not allow for a 
reasonable match to our observations.

The spatial resolution of our observations (0.002 pc) is roughly the 
thickness of the H$_2$ emission region, according to our best-fit model.  
Thus, 
we were not able to spatially resolve the temperature structure of the PDR, 
and our observed width and steepness of the bright ridge is due primarily to 
inclination effects.   
Line intensities and ratios for extended emission behind the main H$_2$ ridge 
can be explained if 1/cos($\theta$) decreases as one moves further into the 
cloud.  This geometry would agree with \citet{hogerheijde}.

Acknowledgements:  
We would like to thank Tommy Greathouse for support of the observing,
Paul van der Werf for providing his results in electronic form, and
David Hollenbach along with our anonymous referee for helpful comments.
This work was supported by NSF grant AST-0205518.  
BTD was supported in part by NSF grant AST-9988126 and in part by
grants from the W.M.~Keck Foundation and the Monell Foundation.
MJR was supported by
NSF grant AST-0307497 and NASA grant NNG04GG92G.  TEXES was built 
and operated by grants from the National Science Foundation and the Texas 
Advanced Research Program.

\clearpage



\clearpage
%%%%%%%%%%%%%%%%% Table 1 %%%%%%%%%%%%%%%%%%%%%%%%%%%%%%%
\begin{deluxetable}{lllccc}
\tablecolumns{4}
%\footnotesize
\tablecaption{Line Data}
\tablewidth{0pt}
\tablehead{
\colhead{Line}                  &
\colhead{Wavelength\tablenotemark{a}}             &
\colhead{E$_{upper}$/k\tablenotemark{b}}            &
\colhead{A\tablenotemark{c}}          \\
\colhead{}                      &
\colhead{$\mu$m}                   &
\colhead{K}                      &
\colhead{s$^{-1}$}     \\                 
}
\startdata
0-0~S(1)      &  17.035  &   1015 & $4.76 \times 10^{-10}$ \\
0-0~S(2)      &  12.279  &   1682 & $2.75 \times 10^{-9}$ \\
0-0~S(4)      &  8.0258  &  3476 & $2.64 \times 10^{-8}$ \\
1-0~S(1)      &  2.1218  &  6955 & $3.47 \times 10^{-7}$ \\
\enddata
\tablenotetext{a}{\citet{jennings}}
\tablenotetext{b}{\citet{mandy}}
\tablenotetext{c}{\citet{wolniewicz}}
\label{lines}
\end{deluxetable}

%%%%%%%%%%%%%%%%%%%%%%%%%Table 2%%%%%%%%%%%%%%%%%%%5
\begin{deluxetable}{rrrrrrr}
\tablecolumns{6}
%\footnotesize
\tablecaption{Observations}
\tablewidth{0pt}
\tablehead{
\colhead{Line}                  &
\colhead{Frequency\tablenotemark{a}}             &
\colhead{slit length}            &
\colhead{slit width}          &
\colhead{Resolving Power}                 &
\colhead{t$_{on}$\tablenotemark{b}}          &
\colhead{rms noise}            \\
\colhead{}                      &
\colhead{cm$^{-1}$}                   &
\colhead{}                     &
\colhead{}                      &
\colhead{}                      &
\colhead{s}                      &
\colhead{}                       \\
}
\startdata
0-0~S(1) & 587.032  & 10.5\arcsec & 2.0\arcsec & 75000 & 160  & 0.32\tablenotemark{c} \\
0-0~S(2) & 814.425  & 7.35\arcsec & 1.4\arcsec & 87000 & 80  & 0.22\tablenotemark{c} \\
0-0~S(4) & 1246.098 & 6.3\arcsec  & 1.4\arcsec & 100000 & 520 & 0.18\tablenotemark{d} \\
\enddata
\tablenotetext{a}{from \citet{jennings}}
\tablenotetext{b}{The average total integration time (ON position) per point}
\tablenotetext{c}{Average noise in units of 10$^{-4}$ erg cm$^{-2}$ s$^{-1}$
sr$^{-1}$ for a single position 
in our smoothed ($\sim$2\arcsec resolution) map.}
\tablenotetext{d} {Average noise for the spectrum derived by summing along the
slit and over all S(4) slit positions.}
\label{observations}
\end{deluxetable}

%%%%%%%%%%%%%%%%%%%%%%%%Table 3%%%%%%%%%%%%%%%%
\begin{deluxetable}{rrrrrrrrrr}
\tablecolumns{10}
%\footnotesize
\tablecaption{Line Widths and Velocities}
\tablewidth{0pt}
\tablehead{
\colhead{}                       &
 \multicolumn{3}{c}{0-0~S(1)}    &
 \multicolumn{3}{c}{0-0~S(2)}                    &
 \multicolumn{3}{c}{0-0~S(4)}     \\
\colhead{Area}                  &
\colhead{$\Delta$V$_{1/2}$\tablenotemark{a}}           &
\colhead{$\Delta$V$_{1/2}$\tablenotemark{b}}           &
\colhead{V$_{LSR}$}           &
\colhead{$\Delta$V$_{1/2}$\tablenotemark{a}}           &
\colhead{$\Delta$V$_{1/2}$\tablenotemark{b}}           &
\colhead{V$_{LSR}$}           &
\colhead{$\Delta$V$_{1/2}$\tablenotemark{a}}           &
\colhead{$\Delta$V$_{1/2}$\tablenotemark{b}}           &
\colhead{V$_{LSR}$}             \\
\colhead{}                       &
\multicolumn{2}{c}{km s$^{-1}$}                     &
\colhead{km s$^{-1}$}                     &
\multicolumn{2}{c}{km s$^{-1}$}                     &
\colhead{km s$^{-1}$}                     &
\multicolumn{2}{c}{km s$^{-1}$}                     &
\colhead{km s$^{-1}$}                     \\
}
\startdata
A & 6.0 & 4.5 & 11.3 & 5.7 & 4.5 & 10.9 & 6.5 & 5.7 & 10.7 \\
B & 5.8 & 4.2 & 11.0 & 5.6 & 4.4 & 10.4 & 6.5 & 5.7 & 9.4 \\
C & 4.6 &$<$4.0&11.0 & 5.6 & 4.4 & 11.5 &     &     & \\
D & 6.7 & 5.4 & 11.1 & 5.6 & 4.4 & 10.4 &     &     & \\
E & 6.0 & 4.5 & 10.7 & 6.9 & 6.0 & 10.3 &     &     & \\
F & 7.7 & 6.6 & 11.0 & 6.7 & 5.7 & 10.2 &     &     & \\
\enddata
\tablenotetext{a}{The best-fit gaussian FWHM to the spectra without
deconvolving the instrument profile}
\tablenotetext{b}{The physical linewidth found by deconvolving the instrument 
profile from the measured linewidth, assuming both are gaussian}
\label{velocities}
\end{deluxetable}

%%%%%%%%%%%%%%%Table 4%%%%%%%%%%%%%%%%%%
\begin{deluxetable}{ccccccccc}
\tablecolumns{8}
%\footnotesize
\tablecaption{H$_2$ Intensities, Temperatures and Column Densities}
\tablewidth{0pt}
\tablehead{
\colhead{}                       &
 \multicolumn{4}{c}{Observed Intensities\tablenotemark{a}}     &
\colhead{}                      &
\colhead{}                      &
\colhead{}                      &
\colhead{}  \\
\colhead{Area}                  &
\colhead{0-0~S(1)}           &
\colhead{0-0~S(2)}           &
\colhead{0-0~S(4)}           &
\colhead{1-0~S(1)\tablenotemark{b}}           &
\colhead{T$_\mathrm{exc}$\tablenotemark{c}}             &
\colhead{T$_\mathrm{exc}$\tablenotemark{d}}             &
\colhead{N(H$_2$)\tablenotemark{e}}            \\
\colhead{}                       &
 \multicolumn{4}{c}{$10^{-4}$ erg cm$^{-2}$ s$^{-1}$ sr$^{-1}$}                     &
\colhead{K}                      &
\colhead{K}                     &
\colhead{10$^{20}$ cm$^{-2}$}  \\
}
\startdata
A & $8.5\pm0.3$ & $6.8\pm0.1$ & $4.1\pm0.3$ & 3.6  & 460 & 500 & 9.0 \\
B & $7.7\pm0.3$ & $5.6\pm0.2$ & $5.1\pm0.4$ & 3.4  & 430 & 570 & 8.9 \\
C & $2.0\pm0.2$ & $2.2\pm0.2$ &             & 1.3  & 590 &     & 1.7 \\
D & $1.6\pm0.2$ & $1.8\pm0.1$ &             & 0.94 & 590 &     & 1.3 \\
E & $3.7\pm0.2$ & $4.4\pm0.2$ &             & 0.49 & 630 &     & 3.0 \\
F & $3.0\pm0.2$ & $1.9\pm0.1$ &             &$<$0.2& 390 &     & 4.0 \\
\enddata
\tablenotetext{a}{for explanation of uncertainties see \S 2}
\tablenotetext{b}{from \citet{vanderwerf}}
\tablenotetext{c}{Determined from I[0-0~S(1)]/I[0-0~S(2)] assuming optically thin, thermalized emission with an equilibrium ortho-to-para ratio.}
\tablenotetext{d}{Determined from I[0-0~S(4)]/I[0-0~S(2)] assuming optically thin, thermalized emission}
\tablenotetext{e}{The total H$_2$ column density determined from I[0-0~S(1)] assuming optically thin, thermalized emission at the temperature determined from I[0-0~S(1)]/I[0-0~S(2)]}
\label{data}
\end{deluxetable}

%%%%%%%%%%%%%%%%%%%%%%%%%%%%  Table 5 %%%%%%%%%%%%%%%%%%%%%%%%%%%%%%
\begin{deluxetable}{ll}
\tablecolumns{2}
\tablecaption{Model Parameters}
\tablewidth{0pt}
\tablehead{}
\startdata
$P/k$      &  $8 \times 10^7 {\rm cm}^{-3} {\rm K}$\\
$\chi$   &  $3 \times 10^4$\\
$\cos(\theta)$ & 0.1 \\
$\zeta_{\rm CR}$ & $1 \times 10^{-15} \rm{s}^{-1}$ \\
C/H       & $1.40 \times 10^{-4}$ \\
O/H       & $3.56 \times 10^{-4}$ \\
Si/H      & $1.74 \times 10^{-6}$ \\
Fe/H      & $2.00 \times 10^{-7}$ \\
\enddata
\label{modelin}
\end{deluxetable}

%%%%%%%%%%%%%%%%%%%%%%%%%Table 6 %%%%%%%%%%%%%%%%%%%%%%%%
\begin{deluxetable}{ccc}
\tablecolumns{3}
\tablecaption{Comparison of $T=1000$~K H-H$_2$ vibrational deexcitation rates}
\tablewidth{0pt}
\tablehead{
\colhead{Reference} &
\colhead{$k_{\rm vdexc.}(1,3)$\tablenotemark{a}} &
\colhead{$k_{\rm vdexc.}(2,3)$\tablenotemark{a}} \\
\colhead{} &
\multicolumn{2}{c}{$({\rm cm}^3\,{\rm s}^{-1})$} \\
}
\startdata
\citet{sternberg89} & $5.5\times10^{-10}$ & $7.5\times10^{-10}$\\
\citet{mandy}  & $6.3\times10^{-12}$ & $3.6\times10^{-11}$\\
\citet{lebourlot} & $3.5\times10^{-13}$ & $5.1\times10^{-13}$\\
present work (see text)   & $5.4\times10^{-11}$ & $7.9\times10^{-11}$\\
\enddata
\tablenotetext{a}{For $k_{\rm vdexc.}(v,J)$ we sum over collisional 
transitions to all levels $(v',J')$ with $v'<v$}
\label{collision}
\end{deluxetable}

%%%%%%%%%%%%%%%%%%%%%%%%%%%% Table 7 %%%%%%%%%%%%%%%%%%%%%%
\begin{deluxetable}{lrrrr}
\tablecolumns{5}
\tablecaption{Model Line Intensities}
\tablewidth{0pt}
\tablehead{
\colhead{Line} &
\colhead{Wavelength}  &
\colhead{Model\tablenotemark{a}} &
\colhead{Observation} &
\colhead{Model/Obs.}  \\
\colhead{} &
\colhead{$\mu$m} &
 \multicolumn{2}{c}{$10^{-3}$ erg cm$^{-2}$ s$^{-1}$ sr$^{-1}$}      &
\colhead{} \\
}
\startdata
H$_2$0-0~S(1) & 17.035   & \ 0.836   & \ 0.85\tablenotemark{b}  &  0.98 \\
H$_2$0-0~S(2) & 12.279   & \ 0.557   & \ 0.68\tablenotemark{b}  &  0.82 \\
H$_2$0-0~S(4)  & 8.0258  & \ 0.439   & \ 0.41\tablenotemark{b} &  1.07 \\
H$_2$1-0~S(1)  & 2.1218  & \ 0.382  & \ 0.36\tablenotemark{c} &  1.06 \\
H$_2$2-1~S(1)  & 2.2477  & \ 0.089  & \ 0.090\tablenotemark{d}&  0.99 \\
SiII & 34.81            & \ 17.4   & \ 7.1\tablenotemark{e}  & 2.1  \\
OI & 63.184             & \ 73.7 & 55.\tablenotemark{f}  &   \\
OI & 145.53             & \ 16.6   & \ 3.\tablenotemark{g}  &   \\
CII & 157.74            & \ 17.2   & \ 4.\tablenotemark{h}  &   \\
\enddata
\tablenotetext{a}{Model 1 (see Table \ref{modelin})}
\tablenotetext{b}{this paper: position A}
\tablenotetext{c}{\citet{vanderwerf}}
\tablenotetext{d}{2-1~S(1)/1-0~S(1) from \citet{usuda_etal_1996}}
\tablenotetext{e}{\citet{stacey95}, 9.2$\arcsec$ beam}
\tablenotetext{f}{\citet{herrmann}, 22$\arcsec$ beam}
\tablenotetext{g}{\citet{herrmann}, 50$\arcsec$ beam}
\tablenotetext{h}{\citet{herrmann}, 55$\arcsec$ beam}
\label{modelout}
\end{deluxetable}


\clearpage
\begin{thebibliography}{}

    \bibitem[Bakes \& Tielens(1994)]{bakes}
Bakes, E.L.O., \& Tielens, A.G.G.M. 1994, ApJ, 427, 822
    \bibitem[Batrla \& Wilson(2003)]{batrla}
Batrla, W., \& Wilson, T.L. 2003, A\&A, 408, 231
%Kinetic temperatures in obar from NH3 inversion lines. Find Tk=150 in mol. gas
    \bibitem[Bertoldi et al.(2000)]{bertoldiiau}
 Bertoldi, F., et al., 2000, IAU Symposium 197, 191
% PDRs and Shocks,  Includes 2023 data and better S140
    \bibitem[Bertoldi \& Draine(1996)]{bertoldidraine}
Bertoldi, F., \& Draine, B.T., 1996, ApJ, 459, 222
% non-stationary pdrs
    \bibitem[Black \& van Dishoeck(1987)]{black}
Black, J.H., \& van Dishoeck, E.F. 1987, ApJ, 332, 412
%Fluorescent excitation of H_2.  The differences between Fluorescent and
% collisional excitation: examine the bar too
     \bibitem[Bohlin, Savage \& Drake(1978)]{bohlin}
Bohlin, R.C., Savage, B.D., \& Drake, J.F. 1978, ApJ, 224, 132
%    \bibitem[Boreiko, Betz, \& Zmuidzinas(1988)]{boreiko}
%Boreiko, R.T., Betz, A.L., \& Zmuidzinas, J. 1988, ApJ, 325, L47
%Original CII work in Orion
    \bibitem[Bregman et al.(1994)]{bregman}
Bregman, J., Larson, K., Rank, D., \& Temi, P. 1994, ApJ, 423, 326
% 3.3 and mid-IR PAH feature mapping and PDR comparison
    \bibitem[Burton, Hollenbach, \& Tielens(1990)]{burton90}
Burton, M.G., Hollenbach, D.J., \& Tielens, A.G.G.M. 1990, ApJ,
    365, 620
%Updates of original Tielens and Hollenbach models, focusing on CO and H2
%O and C+ observations well explained by a clumpy medium.  Nice section on obar
%    \bibitem[Burton, Hollenbach \& Tielens(1992)]{burton92}
%Burton, M.G., Hollenbach, D.J., \& Tielens, A.G.G.M. 1992, ApJ, 399, 563
% compare H2 lines in PDRs and shocks
    \bibitem[Devost et al.(2004)]{devost}
Devost, D., et al. 2004, ApJS, 154, 242
    \bibitem[Draine(2003)]{draine2003a}
Draine, B.T. 2003, ApJ, 598, 1017
    \bibitem[Draine \& Bertoldi(1996)]{draineapj}
Draine, B.T. \& Bertoldi, F. 1996, ApJ, 468, 269
\bibitem[Draine \& Bertoldi(1999)]{draineiso} Draine, B.~T.~\& 
Bertoldi, F.\ 1999, ESA SP-427: The Universe as Seen by ISO, 553
% Bring up the too much hot gas problem and possible additional heating
% mechanisms.  Use Timmermann plots and data
    \bibitem[Draine, Roberge, \& Dalgarno(1983)]{draineshock}
Draine, B.T., Roberge, W.G., \& Dalgarno, A. 1983, ApJ, 264, 485
%MHD shockwaves in molecular clouds
    \bibitem[Felli et al.(1993)]{felli}
Felli, M., Churchwell, E., Wilson, T.L., \& Taylor, G.B. 1993, A\&AS, 98, 137
    \bibitem[Fitzpatrick(1999)]{fitzpatrick99}
Fitzpatrick, E.L. 1999, PASP, 111, 63
    \bibitem[Fuente et al.(1999)]{fuente}
Fuente, A., Mart\'{i}n-Pintado, J., Rodr\'{i}guez-Fern\'{a}ndez, N.J.,
    Rodr\'{i}guez-Franco, A., de~Vincente, P., \& Kunze, D. 1999,
    ApJ, 518, L45
% ISO H2 0-0 S(0), S(1), S(2), S(3), S(4), S(5), S(6) lines toward NGC 7023
% T=300-700K, find O/P of 1.5-2 is needed to explain observations
%   \bibitem[Goldshmidt \& Sternberg(1995)]{goldshmidt}
%  Goldshmidt, O. \& Sternberg, A., 1995, ApJ, 439, 256
% Time-dependent H2 Fluorescence
    \bibitem[Gordon(2004)]{gordon2004}
Gordon, K.D. 2004, in Astrophysics of Dust, 
ed. A.N. Witt, G.C. Clayton, \& B.T. Draine,
ASP Conference Series, 309, 77

    \bibitem[Habart et al.(2004)]{habart} Habart, E., Boulanger, 
F., Verstraete, L., Walmsley, C.~M., \& Pineau des For{\^ e}ts, G.\ 2004, 
\aap, 414, 531
    \bibitem[Habing et al.(1968)]{habing}
Habing, H.J. 1968, BAN, 19, 421
    \bibitem[Hasegawa et al.(1987)]{hasegawa}
Hasegawa, T., Gatley, I., Garden, R.P., Brand, P.W.J.L., Ohishi, M.,
   Hayashi, M., \& Kaifu, N. 1987, ApJ, 318, 77
%    \bibitem[Hayashi et al.(1985)]{hayashi}
%Hayashi, M., Hasegawa, T., Gatley, I., Garden, R., Kaifu, N. 1985, MNRAS, 215, 31
%2-1 and 1-0~S(1) in the bar
    \bibitem[Herrmann et al.(1997)]{herrmann}
Herrmann, F., Madden, S.D., Nikola, T., Poglitsch, A., Timmermann, R.,
    Geis, N., Townes, C.H., \& Stacey, G.J. 1997, ApJ, 481, 343
% C+ and OI in Orion A
    \bibitem[Hogerheijde, Jansen, \& van Dishoeck(1995)]{hogerheijde}
Hogerheijde, M.R., Jansen, D.J., \& van Dishoeck, E.F. 1995, A\&A, 294, 792
% Molecular CO etc. lines...interesting theory of bar structure
%    \bibitem[Hollenbach, \& Natta(1995)]{hollenbachnatta}
%Hollenbach, D.J., \& Natta, A. 1995, ApJ, 455, 133
% Time dependent PDR
    \bibitem[Hollenbach \& Tielens(1999)]{hollenbach}
Hollenbach, D.J., \& Tielens, A.G.G.M. 1999, Rev. Mod. Phys., 71, 173
% Basic review of PDRs
    \bibitem[Hoogerwerf, de Bruijne, \& de Zeeuw(2000)]{hoogerwerf} 
Hoogerwerf, R., de Bruijne, J.~H.~J., \& de Zeeuw, P.~T.\ 2000, \apjl, 544, 
L133 
%distance to Orion
    \bibitem[Howe et al.(1991)]{howe}
Howe, J. E., Jaffe, D.T., Genzel, R. \& Stacey, G.J. 1991, ApJ, 373, 158
    \bibitem[Jansen et al.(1995)]{jansen}
Jansen, D.J., van Dishoeck, E.F., Black, J.H., Spaans, M., \& Sosin, C.
 1995, A\&A, 302, 223
%btd 04.12.26
     \bibitem[Jenkins(2004)]{Jenkins_2004}
Jenkins, E. B. 2004, in Origin and Evolution of the Elements,
     Carnegie Observatories Astrophysics Series
     (Cambridge: Cambridge Univ.\ Press),
     ed. A.~McWilliam and M.~Rauch, 339
%-------
% Modest column density simple PDR
    \bibitem[Jennings, Bragg, \& Brault(1984)]{jennings}
Jennings, D.E., Bragg, S.L, \& Brault, J.W. 1984, ApJ, 282, L85
% Contains wavenumbers of H_2 lines used in paper
     \bibitem[Jura(1975)]{jura}
Jura, M. 1975, ApJ, 197, 575
    \bibitem[Kaufman \& Neufeld(1996)]{kaufman}
Kaufman, M.J., \& Neufeld, D.A. 1996, ApJ, 456, 611
%FIR Water emission from MHD Shock waves
%    \bibitem[K\"oster, St\"orzer, \& Stutzki(1994)]{koster}
%K\"oster, B., St\"{o}rzer, H., Stutzki, J., \& Sternberg, A.
%    1994, A\&A, 284, 545
% CO models, finite cloud size, illuminated from both sides...find substantial 
% H2 cooling component for some densities/FUV fields.
    \bibitem[Lacy et al.(2002)]{lacy}
Lacy, J.H., Richter, M.J., Greathouse, T.K., Jaffe, D.T., \& Zhu, Q. 2002,
    PASP, 114, 153
% TEXES paper
%    \bibitem[Larsson et al.(2003)]{larsson} 
%Larsson, B. et al. 2003, A\&A, 402, L69.
%First NH3 detection in the bar
\bibitem[Le Bourlot et al.(1999)]{lebourlot} Le Bourlot, J., 
Pineau des For{\^ e}ts, G., \& Flower, D.~R.\ 1999, \mnras, 305, 802
% Goes into Cooling by H_2, computes BAD critical densities...
\bibitem[Lemaire et al.(1999)]{lemaire} Lemaire, J.~L., Field, 
D., Maillard, J.~P., Des For{\^ e}ts, G.~P., Falgarone, E., Pijpers, F.~P., 
Gerin, M., \& Rostas, F.\ 1999, \aap, 349, 253 
    \bibitem[Lis \& Schilke(2003)]{lis} 
Lis, D.C. \& Schilke, P. 2003, ApJ, 597, L145
% H13CN and H13CO+ observations of the bar.  Observe at least 10 different 
% molecular condensations. find n_clump of 6E6.  N high enough for collapse to 
% be triggered.  observed H13CN clump parameters consitent with pressure 
% confined clump models.
    \bibitem[Lis et al.(1998)]{lis98}
Lis, D.C., Serabyn, E., Keene, J., Dowell, C.D., Benford, D.J.,
   Phillips, T.G., Hunter, T.R., Wang, N. 1998, ApJ, 509, 299
%    \bibitem[Luhman, Engelbracht, \& Luhman(1998)]{kluhman} 
%Luhman, K.L, Engelbracht, C.W., \& Luhman, M.L. 1998, ApJ, 499, 799.
% NIR spec of obar.  argue obar lines from PDR, not shock, also get HeI lines
    \bibitem[Luhman et al.(1997)]{mluhman} 
Luhman, M.L., Jaffe, D.T., Sternberg, A., Herrmann, F., \& Poglitsch, A. 
    1997, ApJ, 482, 298
% H2, OI, CII emission in Orion A and B.  Compare low density fluorescent 
% excitation case to high density collisional excitation case.
    \bibitem[Mandy \& Martin(1993)]{mandy} 
Mandy, M.E. \& Martin P.G. 1993, ApJ, 86, 199
% Where we get our collision rates...
    \bibitem[Marconi et al.(1998)]{marconi} 
Marconi, A., Testi, L., Natta, A., \& Walmsley, C.M. 1998, A\&A, 330, 696 
% long slit spectra in J, H, K. R of 550 to 950
% no evidence for clumps in H2 emission. densities of 3-6E4, and bar tilted 
%by 10 degrees.  use OI to estimate FUV...find 1-3E4 and conclude electron 
% density of 10E4.  H and He Stromgren sphere coincident.
%btd 04.12.26
    \bibitem[McCall et al.(2003)]{McCall_etal_2003}
McCall, B.J., et al.\ 2003, Nature 422, 500
%-------
    \bibitem[Meixner \& Tielens(1993)]{meixner}
Meixner, M. \& Tielens, A.G.G.M. 1993, ApJ, 405, 216
    \bibitem[Parmar, Lacy, \& Achtermann(1991)]{parmar} 
Parmar, P.S., Lacy, J.H., \& Achtermann, J.M. 1991, ApJ, 372, L25
%Par's original H2 paper
    \bibitem[Parmar, Lacy, \& Achtermann(1994)]{parmarshock}
Parmar, P.S., Lacy, J.H., \& Achtermann, J.M. 1994, ApJ, 430, 786
%shocked H2 in Orion BN-KL
    \bibitem[Plume et al.(1999)]{plume} 
Plume, R., Jaffe, D.T., Tatematsu, K. Evans, N.J. II, \& Keene,
     J. 1999, ApJ, 512, 768
% Plume extended CI paper
    \bibitem[Pogge et al.(1992)]{Pogge etal 1992}
Pogge, R.W., Owen, M.J., \& Atwood, B. 1992, ApJ, 399, 147
    \bibitem[Rieke \& Lebofsky(1985)]{rieke}
Rieke, G.H. \& Lebofsky, M.J. 1985, ApJ, 288, 618
%extinction law
    \bibitem[Rosenthal, Bertoldi \& Drapatz(2000)]{rosenthal}
Rosenthal, D., F. Bertoldi, F., \& Drapatz, S. 2000, A\&A, 356, 705
%o/p of 3, gas way too hot for shock models
%Extinction curve for OMC-1, find o/p of 3
%btd 04.12.26
% X-Ray Modeling of Very Young Early-Type Stars in the Orion Trapezium:
    \bibitem[Schulz et al.(2003)]{Schulz_etal_2003}
Schulz, N.S., Canizares, C., Huenemoerder, D., \& Tibbets, K. 2003,
ApJ, 595, 365
%-------
    \bibitem[Simon et al.(1997)]{simon}
Simon, R., Stutzki, J., Sternberg, A., \& Winnewisser, G. 1997, A\&A, 327, L9
%CN and CS in obar. n=2e5, agrees with edge on homogeneous PDR.
%    \bibitem[Sloan et al.(1997)]{sloan} 
%Sloan, G.C., Bregman, J.D., Geballe, T.R., Allamondola, L.J., \&
%    Woodward, C.E. 1997, ApJ, 474, 735.
    \bibitem[Spaans(1996)]{spaans}
Spaans, M. 1996, A\&A, 307, 271
% long slit spectra 3 micron PAH spectrum.  Find 2 regions of excess emission
% 10'' and 25'' from the ionization front.  
%    \bibitem[Stacey et al.(1993)]{stacey} 
%Stacey, G.J., Jaffe, D.T., Geis, N., Genzel, R., Harris, A.I., 
%    Poglitsch, A., Stutzki, J, \& Townes, C.H. 1993, ApJ, 404, 219.
% CII mapping of Orion
    \bibitem[Stacey et al.(1995)]{stacey95} Stacey, G.~J., Gull, 
G.~E., Hayward, T.~L., Latvakoski, H., \& Peng, L.\ 1995, ASP Conf.~Ser.~ 
73: From Gas to Stars to Dust, 215 
% KWIC imaging of Orion
%    \bibitem[Sternberg(1988)]{sternberg88} 
%Sternberg, A. 1988, ApJ, 332, 400.
% Infrared response of H_2 to FUV: A scaling law. For low n, cold gas.
    \bibitem[Sternberg \& Dalgarno(1989)]{sternberg89} 
Sternberg, A., \& Dalgarno, A. 1989, ApJ, 338, 197
%Infrared response of H_2 gas to UV rad.: high density regions.  Good figures
% of linestrength, cooling, density, FUV...
    \bibitem[Sternberg \& Dalgarno(1995)]{sternberg95}
Sternberg, A., \& Dalgarno, A. 1995, ApJS, 99, 565
% Chemistry of PDRs in detail
    \bibitem[Stutzki et al(1988)]{stutzki} 
Stutzki, J., Stacey, G.J., Genzel, R., Harris, A.I., Jaffe, D.T., 
    \& Lugten, J.B. 1988, ApJ, 332, 379
% Extended CII from M17 Paper
    \bibitem[Tauber et al.(1995)]{tauber95} Tauber, J.~A., Lis, 
D.~C., Keene, J., Schilke, P., \& Buettgenbach, T.~H.\ 1995, \aap, 297, 567 
    \bibitem[Tauber et al.(1994)]{tauber}
Tauber, J.A., Tielens, A.G.G.M., Meixner, M. \& Goldsmith, P.F. 1994,
    ApJ, 423, 136
    \bibitem[Tedds, Brand, \& Burton(1997)]{tedds}
Tedds, J.A., Brand, P.W.J.L., \& Burton, M.G. 1999, MNRAS, 307, 337
%Shocked gas in the Orion Bullets.
%    \bibitem[Tielens \& Hollenbach(1985a)]{th85a} 
%Tielens, A.G.G.M., \& Hollenbach, D. 1985, ApJ, 291, 722.
% The original PDR model of chemistry and heat balance
    \bibitem[Tielens \& Hollenbach(1985b)]{th85b} 
Tielens, A.G.G.M., \& Hollenbach, D. 1985, ApJ, 291, 747
% Modelling the Orion Bar with 1985A paper
    \bibitem[Tielens et al.(1993)]{tielens} 
Tielens A.G.G.M., Meixner, M.M., van der Werf, P.P., Bregman, J., 
    Tauber, J.A., Stutzki, J., \& Rank, D. 1993, Science, 262, 5130
% contains nice false color figure and various cuts. n = 5 E4
    \bibitem[Timmermann et al.(1996)]{timmermann} 
Timmermann, R., Bertoldi, Wright, C.M., Drapatz, S., 
    Draine, B.T., Haser, L., \& Sternberg, A. 1996, A\&A, 315, L281
% ISO observations of H2 in S140: 
%    \bibitem[Turner, Kirby-Docken, \& Dalgarno(1977)]{turner} 
%Turner, J., Kirby-Docken, K., \& Dalgarno, A. 1977, ApJ, 25, 281.
% H2 A values
    \bibitem[Tokunaga(1984)]{tokunaga} Tokunaga, A.~T.\ 1984, \aj, 
89, 172 
%Beta gem and alpha tau N-band mags.
%btd 04.12.26
    \bibitem[Usuda et al.(1996)]{usuda_etal_1996}
Usuda, T., Sugai, H., Kawabata, H., Inoue, M.Y., Kataza, H., \& Tanaka, M. 1996,ApJ, 464, 818
%-------
    \bibitem[van der Werf et al.(1996)]{vanderwerf} 
van der Werf, P.P., Stutzki, J., Sternberg, A., \& Krabbe, A. 
    1996, A\&A, 313, 633
% H2 1-0~S(1) of entire bar 1.5'' res. as well as 13CO, C18O and CS submm lines
    \bibitem[Walmsley et al.(2000)]{walmsley} 
Walmsley, C.M., Natta, A., Oliva, E., \& Testi, L. 2000, A\&A, 364, 301
% NIR Spectral line maps using SOFI. Conclude bar is a cylinder or filament 
% from OI 1.317 micron line.  Measured CI line ratios suggest higher 
% temperatures in the C+ layer than from measured radio line width.  find CI 
% NIR line correlates well with 1-0~S(1) away from high n high FUV areas.
     \bibitem[Weingartner \& Draine(2001)]{weingartner+draine_2001}
Weingartner, J.C., \& Draine, B.T. 2001, ApJS, 134, 263
%    \bibitem[Wright et al.(1999)]{wright}
%Wright, C.M., van Dishoeck, E.F., Cox, P., Sidher, S.D., \& Kessler, M.F. 
%    1999, ApJ, 515, L29.
%ISO observation of HD in the bar...N(H2)=1.5E22
%\bibitem[Wright(2000)]{wright} Wright, C.\ 2000, Molecular 
%hydrogen in space, Cambridge, UK: Cambridge University Press, 2001.~xix, 
%326 p..~Cambridge contemporary astrophysics.~Edited by F.~Combes, and 
%G.~Pineau des For{\^ e}ts.~ISBN 0521782244, p.189, 189 
\bibitem[Wolniewicz et al.(1998)]{wolniewicz} Wolniewicz, L., 
Simbotin, I., \& Dalgarno, A.\ 1998, \apjs, 115, 293
%H2 Einstein A's
    \bibitem[Wyrowski et al.(1997)]{wyrowski} 
Wyrowski, F., Schilke, P., Hofner, P., \& Walmsley, C.M. 1997, ApJ, 487, L171
% C91alpha and C65alpha observations 10'' and 40'' res. nice overall figures,
% derive T for C+ layer of 1500K , n between 5E4 and 2.5E5 C91alp line 
% basically spatially coexistent with 1-0~S(1) data
    \bibitem[Young Owl et al.(2000)]{youngowl}
Young Owl, R.C., Meixner, M.M., Wolfire, M., Tielens, A.G.G.M., \& Tauber, J.
    2000, ApJ, 540, 886
%HCN and HCO+ Images of the bar...

\end{thebibliography}
\end{document}